\documentclass[12pt]{article}

\usepackage{epsf}
\usepackage{latexsym,euscript}
\usepackage[dvips]{graphicx}
\usepackage{amsmath}
\usepackage{amsfonts}
\usepackage{amssymb, epsfig}
\usepackage{cite}

\usepackage{color,soul}

\DeclareMathOperator{\arcsinh}{arcsinh}

\begin{document}

\begin{center}
{\Large \textbf{Dual formulations of Polyakov loop lattice models}}

\vspace*{0.6cm}
\textbf{O.~Borisenko\footnote{email: oleg@bitp.kiev.ua}, 
V.~Chelnokov\footnote{email: chelnokov@bitp.kiev.ua}, 
S.~Voloshyn\footnote{email: billy.sunburn@gmail.com}}

\vspace*{0.3cm}
{\large \textit{ Bogolyubov Institute for Theoretical
Physics, National Academy of Sciences of Ukraine, 03143 Kyiv, Ukraine}}
\vspace*{0.3cm}

\end{center} 

\begin{abstract}
Dual representations are constructed for non-abelian lattice spin models with $U(N)$ 
and $SU(N)$ symmetry groups, for all $N$ and in any dimension. 
These models are usually related to the effective models 
describing the interaction between Polyakov loops in the strong coupled QCD. The original 
spin degrees of freedom are explicitly integrated out and a dual theory appears to be 
a local theory for the dual integer-valued variables. The construction is performed 
for the partition function and for the most general correlation function. The latter include 
the two-point function corresponding to quark--anti-quark free energy and the $N$-point function 
related to the free energy of a baryon. We consider both pure gauge models 
and models with static fermion determinant for both the staggered and Wilson fermions with 
an arbitrary number of flavours. 
While the Boltzmann weights of such models are complex in the presence of non-zero chemical 
potential the dual Boltzmann weights appear to be strictly positive on admissible configurations. 
An essential part of this work with respect to previous studies is an extension of the dual 
representation to the case of 1) an arbitrary value of the temporal coupling constant in 
the Wilson action and 2) an arbitrary number of flavours of static quark determinants.   
The applications and extensions of the results are discussed in detail. 
In particular, we outline a possible approach to Monte-Carlo simulations of the dual theory, 
to the large N expansion and to the development of a tensor renormalization group. 
\end{abstract}   

\newpage 

\section{Introduction}

Dual representations of lattice gauge theories (LGTs) and classical spin models are a useful non-perturbative tool 
that allows to study many aspects of lattice quantum field theories. In the early days of LGT the dual transformations 
proved very efficient in the studies of the confinement and related problems, especially in the abelian gauge theories \cite{dualu1,mack}. 
Also, dual representations appear to be very efficient for numerical simulations both at zero \cite{zachdiss} and 
at finite temperatures for $U(1)$ LGT \cite{dual_u1_dec}. 
The following years have seen many attempts to extend the duality transformations to non-abelian models using different 
approaches and strategies. 
In the pure gauge case the dual representation can be constructed starting from the plaquette formulation \cite{plrepr,plaq_gauge}. 
Dual variables are introduced as variables conjugate to local Bianchi identities \cite{math_probl,conrady}. 
The dual model appears to be non-local due to the presence of connectors in the Bianchi identities for 
gauge models. An analogue of the plaquette formulation for the principal chiral model is so-called 
link representation \cite{link_spin,link_pt}. In this case one can construct a local dual theory 
for all $U(N)$ and $SU(N)$ principal chiral models \cite{spin_dual}. 
Another approach is based on 1) the character expansion of the Boltzmann 
weight and 2) the integration over link variables using the Clebsch-Gordan expansion \cite{gauge_dual,dual4d}. 
The resulting theory is the local dual theory written in terms of invariant $6j$ symbols. Several attempts to simulate 
this dual version have been undertaken (see Ref.\cite{dual_comp1} and references therein). 
In the opposite case, the strong coupling limit, the $SU(N)$ LGT can be mapped onto 
monomer-dimer and closed baryon loop model \cite{karsch}.

During the last decade the dual representations have been applied to solving, fully or partially, the sign problem 
appearing in the lattice QCD in the presence of non-zero chemical potential and/or non-trivial topological term, 
like the $\theta$-term. While it is still too early to say unambiguously if this approach can solve the sign problem 
in QCD, some advances in simpler models are encouraging. For example, the dual form of the massless two-dimensional 
$U(1)$ LGT with one or two flavours of staggered fermions is free of the sign problem \cite{2dqed}. 
The same was proven in the strong coupling limit of the scalar QCD with one, two or three scalar flavours \cite{scalar_qcd}.  

In general, there are two strategies attempting to construct positive Boltzmann weight for QCD or QCD-like theory 
at finite density. The first one relies on full integration over original degrees of freedom, {\it i.e.} gauge and fermion 
fields. The form of the final result strongly depends on the method of integration \cite{nlink_action}, \cite{su2_abc}, 
\cite{su3_abc}, \cite{un_dual}, \cite{unger19}. We do not discuss details of this approach here because in this paper 
we use a second strategy. It consists, first, in construction of an effective model for gauge loops winding around the lattice 
in the temporal direction, {\it i.e.} for Polyakov loops. Only in the second step, the integration over Polyakov loops 
is accomplished. This strategy was successfully applied for the $SU(3)$ Polyakov loop model in the strong coupling region 
for both temporal and spatial couplings of the Wilson action and in the heavy quark regime \cite{spin_flux1}, 
\cite{spin_flux2}, \cite{philipsen12}. Similar results for $U(N)$ models in the same approximations have been presented 
in \cite{un_dual}. More discussion on the effective Polyakov loop models can be found in Refs.\cite{philipsen11}, \cite{philipsen19}.

In this paper we calculate the dual representations for two Polyakov loop models. 
The Boltzmann weight of the first model is the same as the weight studied in \cite{spin_flux1}. 
We extend the results of \cite{spin_flux1} in several directions. First, our calculations are done for all 
values of $N$ and in any dimension. Second, we consider the full static quark determinant with an arbitrary number 
of staggered or Wilson fermion flavours of different masses and chemical potentials. Finally, the result is given 
for the most general correlation function. These include, as particular cases, the partition function, 
two-point function related to the free energy of quark--anti-quark pair and for $N$-point function which gives 
the free energy of a baryon state. 
The Boltzmann weight of the second model is defined for all values of the temporal coupling constant, so that 
the strong coupling limit is imposed only with respect to the spatial coupling. Again, we treat all $SU(N)$ models 
with an arbitrary number of static quark flavours and compute both the partition and correlation functions. 
We shall also explain how the results obtained can be easily transformed into results for $U(N)$ and $Z(N)$ 
Polyakov loop models. 
Boltzmann weights of all dual representations are non-negative, therefore our formulation could be used for 
Monte-Carlo simulations of the models at finite baryon or other chemical potentials.

This paper is organized as follows. 
In Sect.~2 we define the Polyakov loop models and introduce our notations.  
In Sect.~3 we derive dual representations for spin models in the strong coupling region of the temporal coupling constant. 
In Sect.~4 the result is extended to the arbitrary values of the temporal coupling. 
The possible applications and perspectives are discussed in Sect.~5. 
In the Appendix we explain all definitions and our notations related to the group representation theory. 
Also, we evaluate all group integrals appearing in the main text.

\section{Polyakov loop models}

We work on a $d$-dimensional hypercubic lattice $\Lambda = L^d$
with linear extension $L$  and a unit lattice spacing. 
$\vec{x}\equiv x=(x_1,...,x_d)$, $x_i\in [0,L-1]$ denotes the site of the lattice, 
$l=(\vec{x},\nu)$ is the lattice link in the $\nu$-direction and $p=(\vec{x},\mu<\nu)$ is 
the plaquette in the $(\mu,\nu)$-plane. $e_{\nu}$ is a unit vector in the direction $\nu$. 
Periodic boundary conditions are imposed in all directions. 
Let $G=U(N), SU(N)$; $U(x)\in G$, and $dU$ denotes the (reduced) Haar measure
on $G$. ${\rm Tr}U$ will denote the fundamental character of $G$. The character of the irreducible 
representation $\lambda$ will be denoted by $s_{\lambda}(U)$. The dimension of the representation is 
$s_{\lambda}(I)$. 

In this paper we shall study some spin models on $G$ with a local interaction in the external field 
and whose degrees of freedom are labelled by $s_{\lambda}(U)$. These models describe an effective 
interaction between Polyakov loops in $(d+1)$-dimensional LGT with $N_f$ flavours 
of static quarks at finite temperature and non-zero quark chemical potential $\mu$. 
The general form of the partition function of the models is given by  
\begin{eqnarray} 
\label{PF_spindef}
Z_{\Lambda}(\beta,m,\mu;N,N_f)  \equiv  Z \ = \ 
\int \ \prod_x \ dU(x) \prod_{x,\nu} \ B_g(\beta) \ 
\prod_x \ \prod_{f=1}^{N_f} \ B_q(m_f,\mu_f) \ . 
\end{eqnarray}
On an anisotropic lattice and in the limit of vanishing spatial gauge coupling $\beta_s$ one can explicitly 
integrate out all spatial-like fields in any number of dimensions to get the following Boltzmann weight 
describing the Polyakov loop interaction (see, for instance Ref.\cite{caselle} and references therein)
\begin{equation}
B_g(\beta) \ = \ \sum_{\{ \lambda \}} \ D_{\lambda}(\beta) \ s_{\lambda} (U(x)) s_{\lambda} (U^{\dagger}(x+e_{\nu})) \ .
\label{Bgauge}
\end{equation}
The coefficients of  this weight depend on the temporal gauge coupling $\beta_t\equiv\beta$ and can be expressed as 
\begin{eqnarray}
D_{\lambda}(\beta) \ = \ \left ( \frac{C_{\lambda}(\beta)}{s_{\lambda}(I) C_{0}(\beta)} \right )^{N_t} \ , \
C_{\lambda}(\beta) \ = \ \sum_{k=-\infty}^{\infty} \ {\rm det} I_{\lambda_i - i + j + k}(\beta)_{1\leq i,j \leq N} \ . 
\label{D_coeff}
\end{eqnarray}
Here $I_n(x)$ is the modified Bessel function and $N_t$ is the lattice size in the temporal direction. 
In the strong-coupling region $\beta\ll 1$, the leading contribution comes from the fundamental character with 
coefficient $D_F(\beta)$, therefore the whole Boltzmann weight is approximated as 
\begin{equation}
B_g(\beta) \ = \  \exp \left [ \beta_{eff} \ {\rm Re}{\rm Tr}U(x){\rm Tr}U^{\dagger}(x+e_{\nu}) \right ] \ , \ 
\beta_{eff} \ = 2 D_F(\beta) \ . 
\label{Bgauge_strcpl}
\end{equation}
The Boltzmann weight corresponding to the contribution of static staggered fermions can be presented, 
for $N_t$ even, as 
\begin{equation}
B_q(m_f,\mu_f) \ = \ B_{st}(h_+^f,h_-^f) \ = \ 
A_{st} \ {\rm det} \left [ 1 + h_+^f U(x) \right ] \ {\rm det} \left [ 1 + h_-^f U^{\dagger}(x) \right ] \ ,
\label{Zf_stag}
\end{equation}
where the determinant is taken over group indices and 
\begin{equation}
h_{\pm}^f \ = \ e^{-(\arcsinh m_f \mp \mu_f) N_t}  \ . 
\label{hpm_stag}
\end{equation}
The similar formula for static Wilson fermions reads 
\begin{equation}
B_q(m_f,\mu_f) \ = \ B_{w}(h_+^f,h_-^f) \ = \ 
A_w \ {\rm det} \left [ 1 + h_+^f U(x) \right ]^2 \ 
{\rm det} \left [ 1 + h_-^f U^{\dagger}(x) \right ]^2 \ . 
\label{Zf_wilson}
\end{equation}
In this case one has 
\begin{equation} 
h_{\pm}^f \ = \  \left ( 2\kappa_f \ e^{\pm \mu_f} \right )^{N_t} \ , 
\ \kappa_f \ = \ \frac{1}{2m_f+6 +2\cosh\mu_f} \ .
\label{hpm_wilson}
\end{equation} 
The unessential constants
\begin{equation}
A_{st} \ = \ e^{2 N N_t \arcsinh m_f} \ , \ A_w \ = \  (2\kappa_f)^{4 N N_t} 
\label{AstAw}
\end{equation}
will be omitted in the following. 

When $m_f\gg | \mu_f |$ or $\kappa_f\ll e^{\pm \mu_f}$ one usually replaces these exact expressions 
with their approximation 
\begin{equation}
\prod_{f=1}^{N_f} B_q(m_f,\mu_f) \ \approx \ B(h_+,h_-) \ = \ 
\exp \left [  h_+ {\rm {Tr}}U(x) + h_- {\rm {Tr}}U^{\dagger}(x)  \right ] \ ,  
\label{Zf_stag_massive}
\end{equation}
where $h_{\pm}=s \sum_f h_{\pm}^f$, $s = 1$ for the staggered and $s = 2$ for the Wilson fermions. 
The Boltzmann weight of all these models is complex if $\mu_f\ne 0$ or, in general, if $h_+^f\ne h_-^f$. 
In what follows we assume $h_{\pm}^f$ in (\ref{hpm_stag}), (\ref{hpm_wilson}) and (\ref{Zf_stag_massive}) are arbitrary 
complex-valued variables. If $h_{\pm}^f$ are positive, the obtained dual weight is positive, too.

\section{Dual of spin models I}

In this section we consider the partition function (\ref{PF_spindef}) with the weight $B_g(\beta)$ 
given by (\ref{Bgauge_strcpl}). The static fermion contribution $B_q(m_f,\mu_f)$ will be taken either in 
its approximate form (\ref{Zf_stag_massive}) or in exact forms (\ref{Zf_stag}),(\ref{Zf_wilson}). 
The former case has been analyzed in \cite{spin_flux1} for $SU(3)$ by making use of an exact parameterization 
of the $SU(3)$ characters and measure. 

Consider the following Taylor expansion of the Boltzmann weight $B_g(\beta)$
\begin{eqnarray}
\label{taylor_spin_interac}
\exp \left [ \beta_{eff} \ {\rm Re}{\rm Tr}U(x){\rm Tr}U^{\dagger}(x+e_{\nu}) \right ] \ = \ 
\sum_{r=-\infty}^{\infty} \ \sum_{s=0}^{\infty} \ \left ( \frac{\beta_{eff}}{2} \right )^{| r | + 2s} \ 
\frac{1}{(s+| r |)!s!}    \\ 
\left ( {\rm Tr}U(x)  {\rm Tr}U^{\dagger}(x+e_{\nu}) \right )^{s+\frac{1}{2}| r | + \frac{1}{2} r} \ 
\left ( {\rm Tr}U^{\dagger}(x)  {\rm Tr}U(x+e_{\nu}) \right )^{s+\frac{1}{2}| r | - \frac{1}{2} r}  \ . \nonumber 
\end{eqnarray} 
For the fermion weight (\ref{Zf_stag_massive}) we use the similar expansion 
\begin{eqnarray}
\label{taylor_spin_extfield}
\exp \left [ h_+ {\rm {Tr}}U(x) + h_- {\rm {Tr}}U^{\dagger}(x)  \right ] \ = \ 
\sum_{k=-\infty}^{\infty} \ \sum_{m=0}^{\infty}  \ \frac{1}{(m+| k |)! m!} \\  
\left ( h_+ \ {\rm Tr}U(x) \right )^{m+\frac{1}{2}| k | + \frac{1}{2} k} \ 
\left ( h_- \ {\rm Tr}U^{\dagger}(x) \right )^{m+\frac{1}{2}| k | - \frac{1}{2} k} \ .  \nonumber 
\end{eqnarray} 
To deal with exact static determinants (\ref{Zf_stag}) and (\ref{Zf_wilson}) we use an expansion of the determinant 
in the Schur functions (Eq.(\ref{fermdet_exp}) in Appendix), which is valid, in such generality, both for the staggered 
and for the Wilson fermions. Notations and some explanations 
regarding this formula are given in Appendix. 
We shall calculate the dual expression for the most general correlation function 
\begin{equation}
\Gamma(\eta(x),\tilde{\eta}(x)) \ = \ \frac{Z(\eta(x),\tilde{\eta}(x))}{Z} \ \equiv \ 
\left \langle  \prod_x \  \left ( {\rm Tr}U(x) \right )^{\eta(x)} \ 
\left ( {\rm Tr}U^{\dagger}(x) \right )^{\tilde{\eta}(x)}  \right \rangle  \ .
\label{corr_func_def}
\end{equation}
The partition function equals $Z(\eta(x),\tilde{\eta}(x))$ for $\eta(x)=\tilde{\eta}(x)=0$. 

In what follows we analyze separately two cases: 1) heavy quark approximation (\ref{taylor_spin_extfield}) 
and 2) exact static determinant (\ref{fermdet_exp}). All formulas below will be given for $SU(N)$ models. 
In the end, we shall explain how one can easily obtain the corresponding dual 
representations for $U(N)$ and $Z(N)$ models using $SU(N)$ results.

\subsection{Heavy quarks}

The original partition function in the presence of sources $\eta(x),\tilde{\eta}(x)$ is given by 
\begin{eqnarray} 
\label{PF_heavy_quark_1}
Z(\eta(x),\tilde{\eta}(x)) \ = \ \int \ \prod_x \ dU(x) \prod_x \  \left ( {\rm Tr}U(x) \right )^{\eta(x)} \ 
\left ( {\rm Tr}U^{\dagger}(x) \right )^{\tilde{\eta}(x)} \\ 
\prod_{x,\nu} \ \exp \left [ \beta_{eff} \ {\rm Re}{\rm Tr}U(x){\rm Tr}U^{\dagger}(x+e_{\nu}) \right ] \ 
\prod_x \exp \left [ h_+ {\rm {Tr}}U(x) + h_- {\rm {Tr}}U^{\dagger}(x)  \right ] \ .   \nonumber 
\end{eqnarray}
Using (\ref{taylor_spin_interac}) and (\ref{taylor_spin_extfield}) it can be written 
after some rearrangement as 
\begin{eqnarray} 
\label{PF_heavy_quark_2}
&&Z(\eta(x),\tilde{\eta}(x)) \ = \ \prod_l 
\left [ \sum_{r(l)=-\infty}^{\infty} \ \sum_{s(l)=0}^{\infty} \ \left ( \frac{\beta_{eff}}{2} \right )^{| r(l) | + 2s(l)} \ 
\frac{1}{(s(l)+| r(l) |)!s(l)!} \right ]  \\ 
&&\times \prod_x \sum_{k(x)=-\infty}^{\infty} \ \sum_{m(x)=0}^{\infty}  \ 
\frac{\left ( h_+ h_- \right )^{m(x)+\frac{1}{2}| k(x) |}}{(m(x)+| k(x) |)! m(x)!} \ 
\left ( \frac{h_+}{h_-} \ \right )^{\frac{1}{2} k(x)} \   
\prod_x \ Q_N(n(x),p(x))  \ .  \nonumber  
\end{eqnarray}
Here $Q_N(n,p)$ is a group integral defined and calculated in Appendix, Eqs.(\ref{Q_def}), (\ref{QSUN})
\begin{equation} 
Q_N(n,p) \ = \  
\sum_{q=-\infty}^{\infty} \delta_{n-p,q N} \  \bar{Q}_{N,q}(j) \ , \ 
\bar{Q}_{N,q}(j) = \sum_{\lambda \vdash j} \ d(\lambda) \ d(\lambda + |q|^N) \ ,  
\label{barQSUN}
\end{equation}
where $j={\rm min}(n,p)$, $d(\lambda)$ is the dimension of the representation $\lambda$ of the symmetric group $S_n$ 
and the notation $\lambda + |q|^N$ is defined in the Appendix after Eq.(\ref{sun_det}). The integers $n(x)$ and $p(x)$ are given by 
\begin{eqnarray}
\label{nx}
&&n(x) =  t(x) + \frac{1}{2} \sum_{\nu=1}^{d} \left ( r_{\nu}(x) - r_{\nu}(x-e_{\nu}) \right ) 
+ \frac{1}{2} k(x) + \eta(x) \ ,   \\ 
\label{px}
&&p(x) =  t(x) - \frac{1}{2} \sum_{\nu=1}^{d} \left ( r_{\nu}(x) - r_{\nu}(x-e_{\nu}) \right ) 
- \frac{1}{2} k(x) + \tilde{\eta}(x) \ ,  \\ 
&&t(x) = \sum_{i=1}^{2d}  \left ( s(l_i) + \frac{1}{2} | r(l_i) | \right ) + m(x) + \frac{1}{2} |k(x)| \ , 
\label{tx}
\end{eqnarray} 
where $l_i, i=1,...,2d$ are $2d$ links attached to a site $x$ and $s(l)=s_{\nu}(x)$, $r(l)=r_{\nu}(x)$. 
The $N$-ality constraint $n-p=qN$ in (\ref{barQSUN}) becomes 
\begin{equation}
\sum_{\nu=1}^d \left ( r_{\nu}(x) - r_{\nu}(x-e_{\nu}) \right ) + k(x) + \eta(x) - \tilde{\eta}(x) - q(x) N \ = \ 0 \ .  
\label{constr_1}
\end{equation}

\subsubsection{Pure gauge theory} 

Strictly speaking, the conventional duality transformations can be carried out only in the pure gauge theory, {\it i.e.} 
when $h_+=h_-=0$ and, hence $m(x)=k(x)=0$. Then, if $j(x)={\rm min}(n(x),p(x))$, the expression (\ref{PF_heavy_quark_2}) takes the form 
\begin{eqnarray} 
&&Z(\eta(x),\tilde{\eta}(x)) =  \sum_{\{ q(x) \}=-\infty}^{\infty} \sum_{\{ r(l) \} =-\infty}^{\infty} \ \sum_{\{ s(l) \} =0}^{\infty} \ 
\prod_x  \delta_{n(x)-p(x),q(x) N}  \nonumber    \\ 
&&\times\prod_l \left [ \left ( \frac{\beta_{eff}}{2} \right )^{| r(l) | + 2s(l)} \ 
\frac{1}{(s(l)+| r(l) |)!s(l)!} \right ]  \ \prod_x \left [ \bar{Q}_{N,q(x)}(j(x))  \right ]  \ ,  
\label{spin_puregauge}
\end{eqnarray}
while the constraint (\ref{constr_1}) reads 
\begin{equation}
n(x)-p(x) \ = \ \sum_{\nu=1}^d \left ( r_{\nu}(x) - r_{\nu}(x-e_{\nu}) \right )  + \eta(x) - \tilde{\eta}(x) \ = \ q(x) N \ .  
\label{constr_puregauge}
\end{equation}
This constraint can be solved in terms of dual variables in any dimension. 
It is important to emphasize that only $Z(N)$ invariant correlation functions are non-vanishing 
due to above constraint. Indeed, taking into account that on the periodic lattice
$\sum_x \sum_{\nu=1}^d \left ( r_{\nu}(x) - r_{\nu}(x-e_{\nu}) \right ) = 0$ one can be assured that 
\begin{equation}
\sum_x \left ( \eta(x) - \tilde{\eta}(x) \right ) \ = \ N S \ , \ S - \rm{integer} \ .  
\label{constr_puregauge_nality}
\end{equation}
Eq.(\ref{constr_puregauge_nality}) implies that only invariant, {\it i.e.} mesonic and baryonic correlators 
of the Polyajov loops are non-vanishing in the absence of the external field (dynamical quarks). 

In the following we consider, for the sake of simplicity, the two-point correlation function, corresponding to 
the free energy of the quark--anti-quark pair and the $N$-point correlation function, corresponding to the $N$-quark 
(or baryon) potential. In the first case the sources are given by 
\begin{equation}
\eta(x) \ = \ \eta(0) \ = \ \eta \delta_{x,0} \ \ , \ \ 
\tilde{\eta}(x) \ = \ \tilde{\eta}(R) \ = \ \tilde{\eta} \delta_{x,R} \ = \ \eta \delta_{x,R} \ . 
\label{twopoint_source}
\end{equation}
In the second case we introduce sources as 
\begin{equation}
\eta(x) \ = \ \eta(x_i) \ = \ \eta \delta_{x,x_i} \ , \ i=1,\cdots,N \ .  
\label{Npoint_source}
\end{equation}
We give below explicit formulas for $d=1,2,3$ which follow from Eqs.(\ref{spin_puregauge}) and (\ref{constr_puregauge}). 

\subsection*{\underline{One-dimensional model}:}

One-dimensional model is especially simple because we get from (\ref{constr_1})
\begin{equation}
r(l) \ = \ r + k(l) N + \eta(l) \ ,  
\label{constr_1d}
\end{equation} 
where $r\in[0,N-1]$ becomes a global variable, $k(l)\in [-\infty,\infty]$ and $\eta(l)=\eta$ for a set 
of links between sites $x=0,x=R$ and $\eta(l)=0$ for links lying outside of the interval $[0,R]$. 
The delta-function in the 1st line of (\ref{spin_puregauge}) is now $\delta_{k(l)-k(l-1),q(x)}$. 
Making a shift in $q(x)$, the partition function with sources can be presented as 
\begin{eqnarray} 
&&Z(\eta(0),\tilde{\eta}(R)) = \sum_{r=0}^{N-1} \ \sum_{\{ k(l) \} =-\infty}^{\infty} \ 
\sum_{\{ s(l) \} =0}^{\infty} \ \prod_x  \left [ \bar{Q}_{N,k(l)-k(l-1)}(j(x))  \right ]  \nonumber    \\ 
&&\times\prod_l \left [ \left ( \frac{\beta_{eff}}{2} \right )^{| r + k(l) N + \eta(l) | + 2s(l)} \ 
\frac{1}{(s(l)+| r + k(l) N + \eta(l) |)!s(l)!} \right ]  \ ,  
\label{PFspin_ld}
\end{eqnarray}
\begin{equation}
j(x) \ = \  \sum_{i=1}^2 \left ( s(l_i)  + \frac{1}{2} \ | r+k(l_i)N | \right ) 
\pm \frac{1}{2} \ (k(l_1)-k(l_2)) N \ , 
\label{jx_ld}
\end{equation} 
where links $l_1,l_2$ have a site $x$ in common. Signs $"+"$ and $"-"$ corresponds to $n(x)$ and $p(x)$, correspondingly. 

\subsection*{\underline{Two-dimensional model}:}

The solution of the constraint (\ref{constr_1}) in the two-dimensional model and in the presence of sources 
for the quark-anti-quark potential is given by the dual variables as  
(sites are placed in the center of original plaquettes, links are dual to links and sites become dual plaquettes)
\begin{equation} 
r(l) \ = \ r(x) - r(x+e_{\nu}) + k(l) N + \eta(l) \ . 
\label{dual_var2d}
\end{equation}
Here, $\eta(l)=\eta$ if $l\in S_R$, where $S_R$ is some path connecting points $0$ and $R$, and $\eta(l)=0$, otherwise. 
The partition function on the dual lattice takes the form 
\begin{eqnarray} 
&&Z(\eta(0),\tilde{\eta}(R)) = \sum_{\{ r(x) \}=0}^{N-1} \ \sum_{\{ k(l) \} =-\infty}^{\infty} \ 
\sum_{\{ s(l) \} =0}^{\infty} \ \prod_p  \left [ \bar{Q}_{N,k(p)}(j(p))  \right ]  \nonumber    \\ 
&&\times\prod_l \left [ \frac{\left ( \frac{\beta_{eff}}{2} \right )^{| r(x) - r(x+e_{\nu}) + k(l) N + \eta(l) | + 2s(l)}}
{(s(l)+|r(x) - r(x+e_{\nu}) + k(l) N + \eta(l) |)!s(l)!} \right ]  \ ,  
\label{PFspin_2d_2point}
\end{eqnarray}
where we have introduced notations 
\begin{eqnarray}
\label{kp_2d}
k(p) \ = \ k(l_1) + k(l_2) - k(l_3) - k(l_4) \ ,  \\ 
j(p)  =  \sum_{i=1}^4 \left ( s(l_i) + \frac{1}{2} \ | \Delta r(x_i) + k(l_i) N | \right ) \pm \frac{1}{2} \ k(p) \ .  
\label{jp_2d}
\end{eqnarray}
Four links $l_i$ form a dual plaquette $p$ with vertices $x_i$, $\Delta r(x_i)=r(x_i)-r(x_i+e_{\nu})$ and 
signs $"\pm"$ correspond to duals of $n(x)$ and $p(x)$ defined in Eqs.(\ref{nx}), (\ref{px}). 

The solution of the constraint (\ref{constr_1}) in the presence of the baryon sources (\ref{Npoint_source}) can be constructed 
as follows. Let us take an arbitrary point $x_0$ and connect all $N$ points $x_i$ with $x_0$ by some path $S_i$ consisting of 
dual links. Introduce dual variables as in (\ref{dual_var2d}), where $\eta(l)=\eta$ if $l\in S_i$ and $\eta(l)=0$, otherwise. 
The $N$-ality constraint (\ref{constr_1}) becomes 
\begin{equation}
k(p) + \eta \delta_{p,p_0} \ = \ q(p) \ , 
\label{Nality_2d}
\end{equation}
where the plaquette $p_0$ is dual to the site $x_0$ and $k(p)$ is the same as in (\ref{kp_2d}). 
Strictly speaking, the solution of the form (\ref{dual_var2d}) is only valid in two dimensions if $N\leq 4$. 
Then one can take all paths $S_i$ consisting of non-intersecting links and solution (\ref{dual_var2d}) holds.  
Though it is not a problem to extend the solution (\ref{dual_var2d}) to arbitrary $N$ we restrict ourselves here 
to the case $N\leq 4$. We thus conclude that the partition function in the presence of such baryon sources is of 
the form (\ref{PFspin_2d_2point}), where one has to substitute $k(p) \to k(p) + \eta \delta_{p,p_0}$.

\subsection*{\underline{Three-dimensional model}:}

In the physically most relevant three dimensional case one obtains the solution of (\ref{constr_1}) 
in the following form 
\begin{equation} 
r(l) \ = \ r(l_1) + r(l_2) - r(l_3) - r(l_4) +  k(p) N + \eta(p) \ \equiv \ r(p) + k(p) N + \eta(p) \ .  
\label{dual_var3d}
\end{equation}
Here, four links $l_i$ form a plaquette $p$ dual to the original link $l$. 
$\eta(p)=\eta$ if $l\in S_R$, where $S_R$ is some path consisting of dual plaquettes and 
connecting points $0$ and $R$, and $\eta(p)=0$, otherwise. 
The partition function on the dual lattice reads 
\begin{eqnarray} 
&&Z(\eta(0),\tilde{\eta}(R)) = \sum_{\{ r(l) \}=0}^{N-1} \ \sum_{\{ k(p) \} =-\infty}^{\infty} \ 
\sum_{\{ s(p) \} =0}^{\infty} \ \prod_c  \left [ \bar{Q}_{N,k(c)}(j(c))  \right ]  \nonumber    \\ 
&&\times\prod_p \left [ \frac{\left ( \frac{\beta_{eff}}{2} \right )^{| r(p) + k(p) N + \eta(p) | + 2s(p)}}
{(s(p)+ | r(p) + k(p) N + \eta(p) |)!s(p)!} \right ]  \ . 
\label{PFspin_3d_2point}
\end{eqnarray}
$\prod_c$ is a product over all cubes of the dual lattice and the notations are used 
\begin{eqnarray}
\label{kc_3d}
k(c) \ = \ k(p_1) + k(p_2) + k(p_3) - k(p_4) - k(p_5) - k(p_6) \ ,  \\ 
j(c)  =  \sum_{i=1}^6 \left ( s(p_i) + \frac{1}{2} \ | r(p_i) + k(p_i) N | \right ) \pm \frac{1}{2} \ k(c) \ .  
\label{jc_3d}
\end{eqnarray}
Six plaquettes $p_i$ form a dual cube $c$  and signs $"\pm"$ correspond to duals of $n(x)$ and $p(x)$ 
defined in Eqs.(\ref{nx}), (\ref{px}). 

Extension of this result to the $N$-point correlation function is done precisely like in two-dimensional theory. 
In particular, if $N\leq 6$ the solution of (\ref{constr_1}) can be taken as in (\ref{dual_var3d}). 
Then, defining paths $S_i, i=1,\cdots,N$ that connect points $x_i$ with some reference point $x_0$ 
(on the dual lattice path $S_i$ is formed out of plaquettes and connects cubes $c_i$ and $c_0$ which are dual 
to the corresponding sites) and introducing sources $\eta(p)=\eta$ on plaquettes belonging to $S_i$ one finds 
that the $N$-point correlation function is described by Eq.(\ref{PFspin_3d_2point}) where one has to take 
the corresponding sources $\eta(p)$ and make the substitution $k(c) \to k(c) + \eta \delta_{c,c_0}$.

We can conclude that all three-dimensional $SU(N)$ spin models are dual to the gauge models whose partition 
function is given by Eq.(\ref{PFspin_3d_2point}) with $\eta(0)=\tilde{\eta}(R)=0$.

\subsubsection{Full theory} 
\label{full_theory}

Here we proceed with the full theory given by Eq.(\ref{PF_heavy_quark_2}). Using $N$-ality constraint (\ref{constr_1}) 
one can sum up over $k(x)$. With the help of notation 
\begin{equation}
r(x) \ = \ \sum_{\nu=1}^d \left ( r_{\nu}(x-e_{\nu}) - r_{\nu}(x) \right ) \ , 
\label{kx_def}
\end{equation}
we obtain after some manipulations the following expression 
\begin{align} 
&Z(\eta(x),\tilde{\eta}(x)) = \sum_{\{ q(x) \}, \{ r(l) \}=-\infty}^{\infty}   
\sum_{\{ m(x) \}, \{ s(l) \} =0}^{\infty} \ 
\prod_x  \left [ e^{\mu (q(x)N + \tilde{\eta}(x) - \eta(x))} \ \bar{Q}_{N,q(x)}(j(x))  \right ] \nonumber  \\ 
&\times \prod_l  \frac{\left ( \frac{\beta_{eff}}{2} \right )^{| r(l) | + 2s(l)}}{(s(l)+| r(l) |)!s(l)!}
\prod_x \frac{h^{2 m(x)+| r(x) + q(x)N + \tilde{\eta}(x) - \eta(x) |}}{(m(x)+ | r(x) + q(x)N + \tilde{\eta}(x) - \eta(x) |)! m(x)!} \ . 
\label{PF_heavy_quark_3}
\end{align}
We used here the property $\sum_x r(x) = 0$ and introduced parametrization 
\begin{equation}
h_{\pm} \ = \ h e^{\pm \mu} \ . 
\label{hpm_h_mu}
\end{equation}
The expression (\ref{PF_heavy_quark_3}) is our final dual representation for $SU(N)$ Polyakov loop models valid for all $N$ and 
in any dimension. The function $\bar{Q}_{N,q(x)}(j(x))$ is defined in Eq.(\ref{barQSUN}) with $j(x)={\rm min}(j_+(x),j_-(x))$
and $j_{\pm}(x)$ is given by 
\begin{eqnarray}
\label{jx_final}
j_{\pm}(x) \ = \ t(x) + \frac{1}{2} \ \left ( \eta(x) + \tilde{\eta}(x) \pm q(x) N \right ) \ ,   \\ 
t(x) \ = \ \sum_{i=1}^{2d} \ \left ( s(l_i) + \frac{1}{2} \ | r(l_i) | \right ) + m(x) + 
\frac{1}{2} \ | r(x) + q(x)N + \tilde{\eta}(x) - \eta(x) | \ .  \nonumber 
\end{eqnarray} 

Some comments are in order:
\begin{itemize}
\item 
As follows from Eq.(\ref{PF_heavy_quark_3}) and exact expression for the function $Q_{N,q}(j)$ given in (\ref{barQSUN}), 
the dual Boltzmann weight is non-negative if $h_+,h_- > 0$ or if $h_+,h_- < 0$. Hence, in this region the dual formulation 
can be used for the numerical simulations of the model with non-vanishing chemical potentials. 
\item 
Most thermodynamical functions and local physical observables, like the energy density, the baryon density, the quark condensate, etc. 
can be easily translated into the dual form by taking the corresponding derivatives with respect to $\beta_{eff}$, $h_{\pm}^f$ 
or $\mu_f$. This amounts to a local shift in a corresponding summation variable and can be presented as an expectation value 
calculated over the dual partition function. 
\item 
The long-distance observables, like two- and $N$-point correlation functions can also be written as an 
expectation values in the dual form. This follows directly from (\ref{PF_heavy_quark_3}). 
\end{itemize}

\subsubsection{$U(N)$ and $Z(N)$ models} 

Here we explain briefly how the general result for $SU(N)$ models can be used to compute 
the corresponding dual representations for $U(N)$ and $Z(N)$ models. The latter is equivalent to vector 
Potts models and can be obtained from $SU(N)$ models by replacing $U(x)$ matrices with their center elements. 
For simplicity we restrict ourselves here to the partition functions, {\it i.e.} $\eta(x)=\tilde{\eta}(x)=0$.

\subsection*{\underline{$U(N)$ model}:} 

As explained in the Appendix, the only term contributing to $U(N)$ group integrals is the term with $q(x)=0$. 
Therefore, from Eq.(\ref{PF_heavy_quark_3}) one gets for the partition function 
\begin{align} 
&Z = \sum_{\{ q(x) \}, \{ r(l) \}=-\infty}^{\infty}   
\sum_{\{ m(x) \}, \{ s(l) \} =0}^{\infty} \ 
\prod_x  \left [ \bar{Q}_{N}(j(x))  \right ] \nonumber  \\ 
&\times \prod_l  \frac{\left ( \frac{\beta_{eff}}{2} \right )^{| r(l) | + 2s(l)}}{(s(l)+| r(l) |)!s(l)!}
\prod_x \frac{h^{2 m(x)+| r(x) |}}{(m(x)+ | r(x) |)! m(x)!} \ , 
\label{PF_heavy_quark_UN}
\end{align}
 \begin{eqnarray}
\label{jx_final_UN}
j(x) \ = \ \sum_{i=1}^{2d} \ \left ( s(l_i) + \frac{1}{2} \ | r(l_i) | \right ) + m(x) + \frac{1}{2} \ | r(x) | \ . 
\end{eqnarray}

\subsection*{\underline{$Z(N)$ model}:} 

Even simpler is the result for $Z(N)$ model. In this case $\bar{Q}_{N,q(x)}(j(x))=1$. 
Taking into account that 
\begin{equation} 
\sum_{s =0}^{\infty} \  \frac{\left ( \frac{x}{2} \right )^{r + 2s}}{(s+r)!s!} \ = \ I_r(\beta) \ , 
\label{sum_bessel}
\end{equation}
where $I_r(x)$ is the modified Bessel function, the partition function appears to be 
\begin{equation} 
Z = \sum_{\{ q(x) \}, \{ r(l) \}=-\infty}^{\infty}   e^{\mu N \sum_x q(x)} \  \prod_l  I_{r(l)}(\beta_{eff}) \ 
\prod_x I_{r(x)+q(x)N}(h) \ . 
\label{PF_heavy_quark_ZN}
\end{equation}

Let us add some more comments here: 
\begin{itemize}
\item 
Clearly, all comments given in the end of Sec.(\ref{full_theory}) remain valid for $U(N)$ and $Z(N)$ models. 
\item 
It follows from (\ref{PF_heavy_quark_UN}) that the partition function and invariant observables 
do not depend on the chemical potential for $U(N)$ models with one fermion flavour.
In case of two flavours, the Boltzmann weight depends only on the difference of chemical potentials $\mu_1-\mu_2$. 
\item 
In the pure gauge case the corresponding representations for $U(N)$ and $Z(N)$ models can be straightforwardly obtained 
from Eqs.(\ref{PFspin_ld}), (\ref{PFspin_2d_2point}) and (\ref{PFspin_3d_2point}). 
\item 
The dual form of the $XY$ model can be calculated either taking $N=1$ in $U(N)$ case or as a limit $N\to\infty$ in $Z(N)$ case. 
{\it E.g.}, in the pure gauge three-dimensional case one recovers the following dual gauge-like form of the $XY$ model 
\begin{eqnarray}
Z_{XY}(\beta)  \ = \ \sum_{\{ r(l) \} =-\infty}^{\infty} \ \prod_l \ I_{r(p)}(\beta) \ , \ 
r(p) = r(l_1) + r(l_2) - r(l_3) - r(l_4) \ . 
\label{u1_dual}
\end{eqnarray} 
\end{itemize}

\subsection{Exact static determinant} 

In this subsection, we compute the dual representation for the theory with the exact static determinant with
an arbitrary number of flavours of the staggered, Eq.(\ref{Zf_stag}), or the Wilson, Eq,(\ref{Zf_wilson}), fermions. 
As in the previous subsection, we shall calculate the dual expression for the most general correlation function 
defined in (\ref{corr_func_def}). 
The original partition function in the presence of sources $\eta(x),\tilde{\eta}(x)$ is given by 
\begin{eqnarray} 
\label{PF_statdet_1}
Z(\eta(x),\tilde{\eta}(x)) \ = \ \int \ \prod_x \ dU(x) \prod_x \  \left ( {\rm Tr}U(x) \right )^{\eta(x)} \ 
\left ( {\rm Tr}U^{\dagger}(x) \right )^{\tilde{\eta}(x)} \\ 
\prod_{x,\nu} \ \exp \left [ \beta_{eff} \ {\rm Re}{\rm Tr}U(x){\rm Tr}U^{\dagger}(x+e_{\nu}) \right ] \ 
\prod_x  \prod_{f=1}^{N_f} \ B_q(m_f,\mu_f)   \ .   \nonumber 
\end{eqnarray}
The gauge part of the Boltzmann weight $B_g(\beta)$ is treated as in the previous subsection using the expansion (\ref{taylor_spin_interac}). 
Substituting this expansion into (\ref{PF_statdet_1}) one gets after some rearrangement 
\begin{eqnarray} 
Z(\eta(x),\tilde{\eta}(x)) &=& \prod_l 
\left [ \sum_{r(l)=-\infty}^{\infty} \ \sum_{s(l)=0}^{\infty} \ \left ( \frac{\beta_{eff}}{2} \right )^{| r(l) | + 2s(l)} \ 
\frac{1}{(s(l)+| r(l) |)!s(l)!} \right ]    \nonumber   \\ 
&\times& \prod_x  R_{N,N_f}(n(x),p(x);m_f,\mu_f) \ .  
\label{PF_statdet_2}
\end{eqnarray}
Here, the function $R_{N,N_f}(r,s;m_f,\mu_f)$ is a group integral defined in Eq.(\ref{Rint_def}) of Appendix. 
The integers $n(x)$ and $p(x)$ are given by 
\begin{eqnarray}
\label{nx_stat}
&&n(x) =  \sum_{i=1}^{2d}  \left ( s(l_i) + \frac{1}{2} | r(l_i) | \right )
+ \frac{1}{2} \sum_{\nu=1}^{d} \left ( r_{\nu}(x) - r_{\nu}(x-e_{\nu}) \right ) 
+ \eta(x) \ ,   \\ 
\label{px_stat}
&&p(x) =  \sum_{i=1}^{2d}  \left ( s(l_i) + \frac{1}{2} | r(l_i) | \right )
- \frac{1}{2} \sum_{\nu=1}^{d} \left ( r_{\nu}(x) - r_{\nu}(x-e_{\nu}) \right ) 
+ \tilde{\eta}(x) \ ,  
\end{eqnarray} 
where $l_i, i=1,...,2d$ are $2d$ links attached to a site $x$. 

To deal simultaneously with staggered and Wilson fermions we use the representation (\ref{fermdet_exp}) for 
the $N_f$-flavour static determinant proven in Appendix. With this representation and making use Eq.(\ref{trace_power}) 
the group integral can be calculated exactly. This is done in Appendix, formulas (\ref{Rint_gen})-(\ref{Rint_res2}). 
Presenting the $N$-ality constraint as 
\begin{equation}
g(x) - f(x) \ = \ q(x) N \ ; \  g(x) = n(x) + |\alpha(x)| \ , \ f(x) = p(x) + |\beta(x)|  \ , 
\label{Nality_stat}
\end{equation}
we write down the final result (\ref{Rint_res2}) in the explicit form  
\begin{align} 
&R_{N,N_f}(n(x),p(x);m_f,\mu_f) = \sum_{q(x)=-\infty}^{\infty} \ 
\sum_{|\alpha(x)| =0}^{c N N_f} \ \sum_{|\beta(x)| =0}^{c N N_f}
\delta_{g(x) - f(x), q(x)N} \nonumber   \\  
\label{Rint_fin}
&\times \ \bar{R}_{N,N_f}(q(x),|\alpha(x)|, |\beta(x)|; H_{\pm})   \  ;  \ \ \ 
\bar{R}_{N,N_f}(q(x),|\alpha(x)|, |\beta(x)|; H_{\pm})   \\ 
&= \sum_{\alpha \vdash |\alpha(x)|} \ \sum_{\beta \vdash |\beta(x)|} \ 
\sum_{\sigma \vdash n(x) + |\alpha(x)|} \ d(\sigma/\alpha) d(\sigma+q^N/\beta) \ 
s_{\alpha^{\prime}}(H_+) s_{\beta^{\prime}}(H_-) \ ,   \nonumber 
\end{align}
where $c=1$ for the staggered and $c=2$ for the Wilson fermions. 
The explicit form of the $N$-ality constraint is
\begin{equation}
g(x) - f(x) \ \equiv \  \sum_{\nu=1}^{d} \left ( r_{\nu}(x) - r_{\nu}(x-e_{\nu}) \right ) 
+ \eta(x) -\tilde{\eta}(x) + |\alpha(x)| -  |\beta(x)| = q(x) N \ . 
\label{Nality_stat_2}
\end{equation}
The variables $H_{\pm}$, depending on $m_f,\mu_f$, and other notations are defined and described in Appendix. 
Combining last expressions with (\ref{PF_statdet_2}) the final result for the partition function 
with arbitrary sources gets the form 
\begin{align} 
\label{PF_statdet_3}
&Z(\eta(x),\tilde{\eta}(x)) = \sum_{\{ q(x) \} =-\infty}^{\infty}  \ \sum_{\{ |\alpha(x)| \}=0}^{c N N_f}  \ 
\sum_{\{ |\beta(x)| \}=0}^{c N N_f} \  
\sum_{\{ r(l) \} = -\infty}^{\infty} \ \sum_{\{ s(l) \}= 0}^{\infty}     \\ 
&\prod_l  \frac{\left ( \frac{\beta_{eff}}{2} \right )^{| r(l) | + 2s(l)}}{(s(l)+| r(l) |)!s(l)!} \ 
\prod_x  \left [ \delta_{g(x) - f(x), q(x)N} \  
\bar{R}_{N,N_f}(q(x),|\alpha(x)|, |\beta(x)|; H_{\pm}) \right ]  \ . \nonumber  
\end{align}

We conclude this subsection with few comments: 
\begin{itemize}
\item 
All factors entering the Boltzmann weight of (\ref{PF_statdet_3}) are positive. This is true 
also for the Schur functions appearing in (\ref{Rint_fin}). Hence, this representation is suitable 
for the numerical simulations of the theory. 

\item 
Both long-distance and local observables can be written as expectation values over the dual partition function. 

\item 
An explicit form of the group integral for one flavour of the staggered fermions is presented in Appendix, Section (\ref{stag_1fl}). 
Even more detailed formula is given there for $N=3$. This presentation can be directly used for the Monte-Carlo simulations. 

\item 
The dual form of $U(N)$ model coincides with Eq.(\ref{PF_statdet_3}) where one should take the only term with $q(x)=0$.
The dual form of $Z(N)$ model is obtained from Eq.(\ref{PF_statdet_3}) by replacing all dimensions  
with unity and omitting sum over $\sigma$ in the definition of $\bar{R}_{N,N_f}$, Eq.(\ref{Rint_fin}). 
\end{itemize}

\section{Dual of spin models II}

Here we investigate the partition function (\ref{PF_spindef}) with the weight $B_g(\beta)$ 
given by (\ref{Bgauge}). The static fermion contribution $B_q(m_f,\mu_f)$ will be taken in 
its exact forms (\ref{Zf_stag}),(\ref{Zf_wilson}). Like in the previous section we calculate 
the dual form both for the partition function and for the most general correlation function. 
In this case such correlations are given by an expectation value of the product of $SU(N)$ 
characters taken in arbitrary representations $\eta(x),\tilde{\eta}(x)$. Precisely, one has 
\begin{eqnarray} 
\label{PF_spindef_II_1}
Z(\eta(x),\tilde{\eta}(x)) \ = \ \int \ \prod_x \ dU(x) 
\prod_x \left [  s_{\eta(x)} (U(x)) s_{\tilde{\eta}(x)} (U^{\dagger}(x)) \right ]  \\ 
\prod_{x,\nu} \left [ \sum_{\{ \lambda \}} \ D_{\lambda}(\beta) \ s_{\lambda} (U(x)) s_{\lambda} (U^{\dagger}(x+e_{\nu})) \right ] \ 
\prod_x \ \prod_{f=1}^{N_f} \ B_q(m_f,\mu_f) \ .  \nonumber  
\end{eqnarray}
The partition function is recovered by taking trivial representations in all lattice sites. 
Exchanging order of the summations and integrations and rearranging product over links in the second 
line of (\ref{PF_spindef_II_1}) we write down the result in the form 
\begin{eqnarray} 
\label{PF_spindef_II_2}
Z(\eta(x),\tilde{\eta}(x)) \ = \ \sum_{\{ \lambda(l) \}} \ 
\prod_{x,\nu} \left [ D_{\lambda(l)}(\beta) \right ] \ \prod_x \ 
H_{N,N_f}^d(g(x),f(x);m_f,\mu_f) \ . 
\end{eqnarray}
The coefficients $D_{\lambda}(\beta)$ are defined in (\ref{D_coeff}) and the function $H_{N,N_f}^d$ is a group integral 
defined in (\ref{Hint_def}) and calculated in Appendix, Eqs.(\ref{Hint_d0})-(\ref{Hint_dn}), where 
\begin{equation} 
g(x) \ = \ (\lambda_{\nu}(x),\eta(x)) \ , \ f(x) \ = \ (\lambda_{\nu}(x-e_{\nu}),\tilde{\eta}(x)) \ .
\label{gx_fx_II}
\end{equation}
In the next subsections we specify this general formula for several important cases.

\subsection{Pure gauge theory} 

In the pure gauge theory the group integral $H_{N,N_f}^d$ simplifies to $G_N^d(\lambda_i,\gamma_i )$ given by Eq.(\ref{Gint_def}). 
The result of the integration is given by Eqs.(\ref{GN1})-(\ref{GN3}). Denoting $r(l)\equiv r_{\nu}(x)=|\lambda(l)|$, 
the sum over all representations $\lambda$ can be written as 
\begin{equation}
\sum_{\lambda} \cdots  \ = \ \sum_{r=0}^{\infty} \ \sum_{\lambda\vdash r} \cdots   \  
\label{sum_repr}
\end{equation}
and the $N$-ality constraints (\ref{nality_GN2}) and (\ref{nality_GN3}) in the presence of sources read
\begin{equation}
r(x) + | \eta(x) | - | \tilde{\eta}(x) | \ = \ q(x) N \ ; \ 
r(x) \ = \ \sum_{\nu=1}^d \ \left ( r_{\nu}(x) - r_{\nu}(x-e_{\nu}) \right ) \ .  
\label{nality_II_gauge}
\end{equation}
As before, we analyze each dimension separately. The result will be given for the partition function and for the correlation 
function of the general form. We shall also explain the particular solution of the $N$-ality constraints 
for the two- and $N$-point correlation functions. The corresponding sources can be taken like in 
Eqs.(\ref{twopoint_source}) and (\ref{Npoint_source}), respectively.

\subsection*{\underline{One-dimensional model}:}

One-dimensional model is exactly solvable. Using orthogonality relation (\ref{GN1}) and expression for 
the Littlewood-Richardson coefficients $C^{\eta}_{\lambda_1 \ \lambda_2}$ (\ref{LR_int}) one finds for the partition function
\begin{eqnarray} 
\label{PF_II_puregauge_1d_1}
Z \ = \ \sum_{\{ \lambda \}} \ \left [ D_{\lambda}(\beta) \right ]^L 
\end{eqnarray}
and for the two-point function
\begin{eqnarray} 
\label{PF_II_puregauge_1d_2}
Z(\eta(0),\tilde{\eta}(R)) \ = \ \sum_{\{ \lambda_1,\lambda_2 \}} \ C^{\eta}_{\lambda_1 \ \lambda_2} \ C^{\tilde{\eta}}_{\lambda_2 \ \lambda_1} \ 
\left [ D_{\lambda_1}(\beta) \right ]^R \ \left [ D_{\lambda_2}(\beta) \right ]^{L-R} \ .
\end{eqnarray}
Summations in the last expressions goes over $SU(N)$ representations 
$\lambda_1\geq\lambda_2\geq \cdots \geq \lambda_{N-1}\geq \lambda_N=0$.

\subsection*{\underline{Two-dimensional model}:}

\begin{figure}[t]
\centerline{{\epsfxsize=7cm \epsfbox{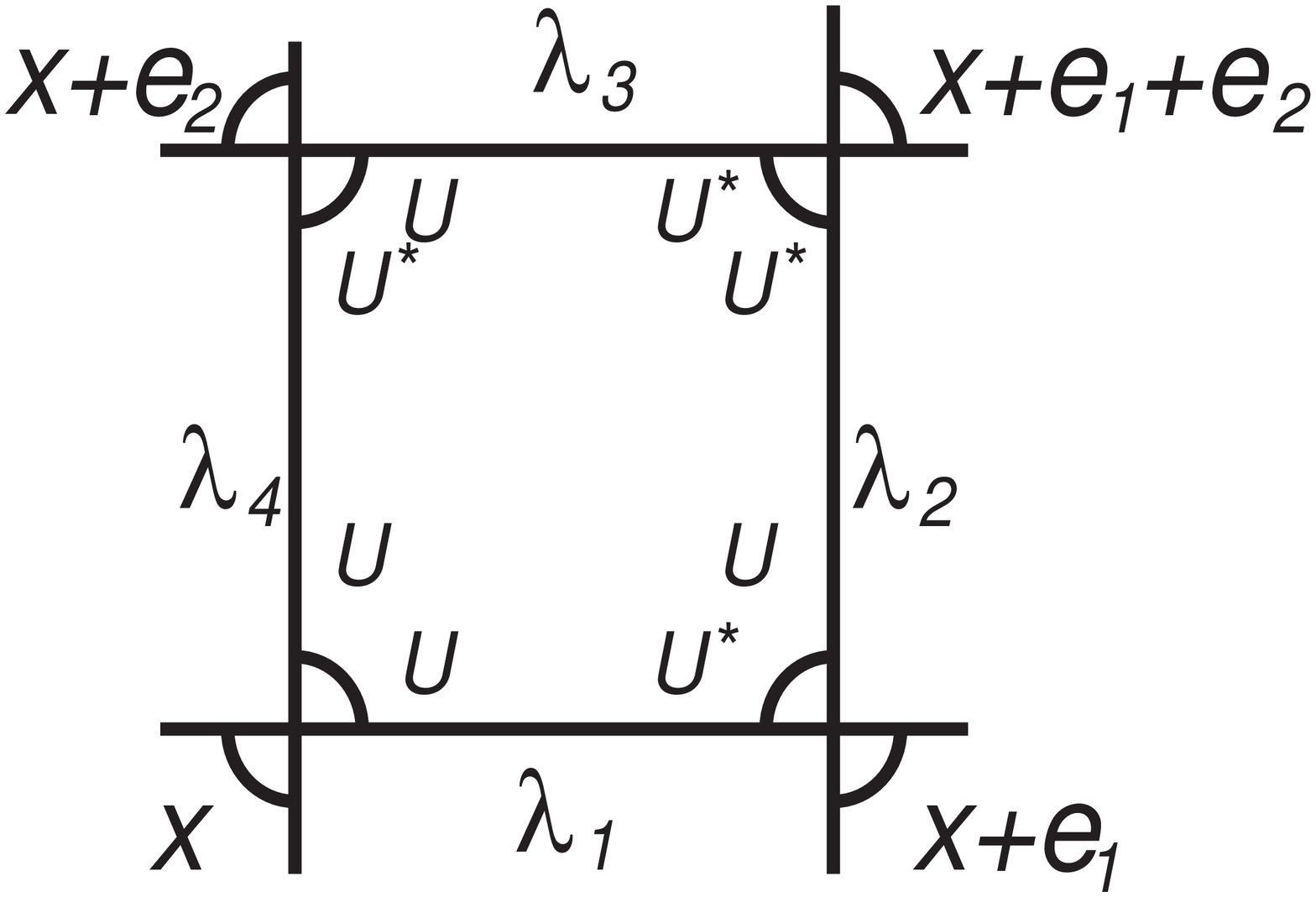}} {\epsfxsize=5cm\epsfbox{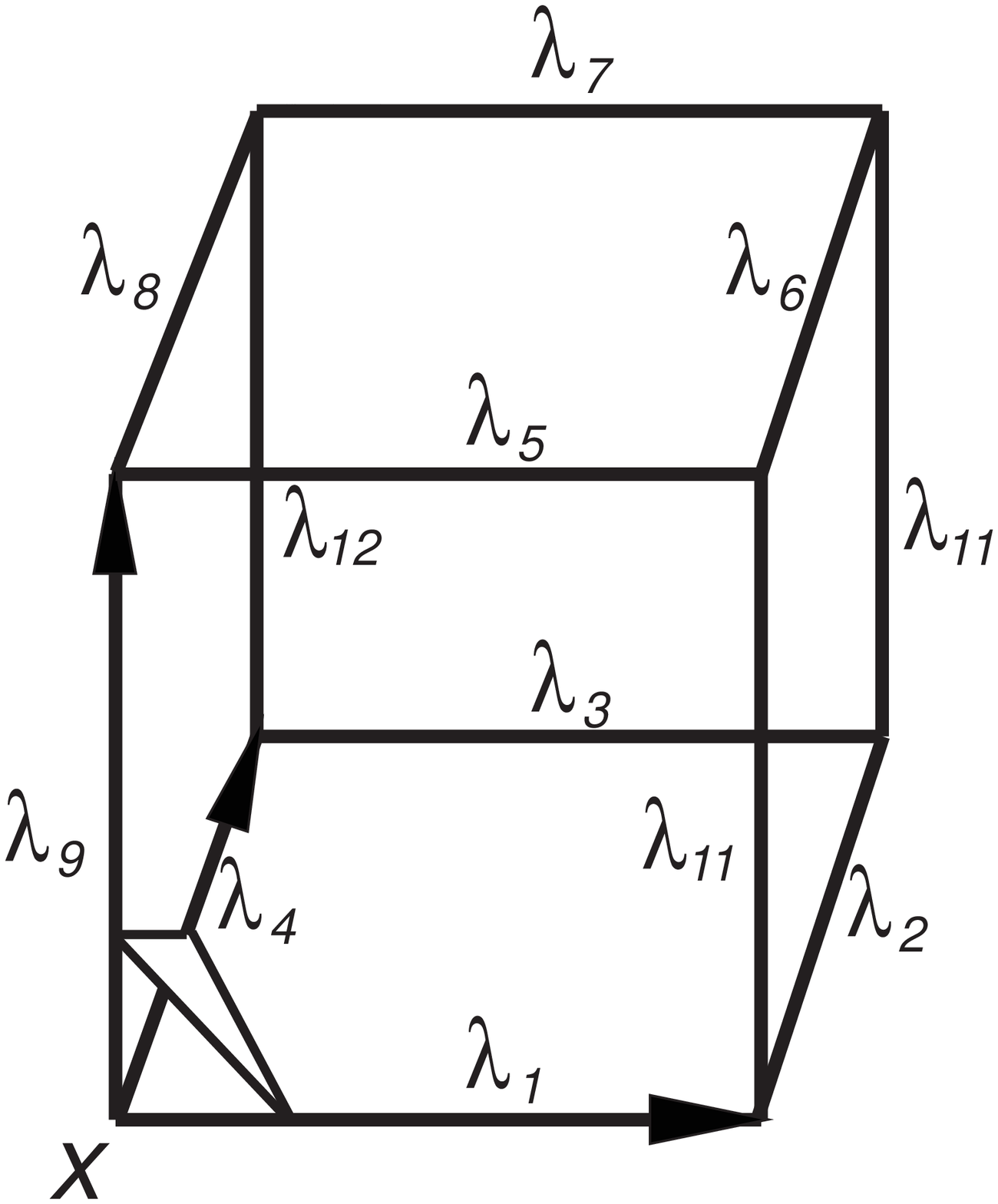}}}
\caption{\label{schur_coupl_comb} The order of coupling of the link representatons in the integrand of one-site group integrals. 
Left: in two-dimensional theory representations are coupled inside every even plaquette of the lattice. 
Right: in three-dimensional theory representations are coupled inside every even cube of the lattice.}
\end{figure}

In two- and three-dimensional cases the final result can be significantly simplified if we multiply the Schur functions 
in the integrand in a special way. Namely, in two dimensions we divide the lattice into a set of even and odd plaquettes and 
couple the Schur functions as shown in the left panel of Fig.(\ref{schur_coupl_comb}). 
In this way the summations over original representations $\lambda(l)$ are factorized inside every even plaquette. 
Let $\bar{\lambda}$ be a representation conjugate to $\lambda$, see Eqs.(\ref{schur_conj_un}), (\ref{schur_conj_sun}). 
Extending the integration result (\ref{GN2}) to the correlation functions and using decomposition (\ref{sum_repr}), 
we obtain 
\begin{eqnarray}
Z(\eta(x),\tilde{\eta}(x)) &=& \sum_{\{ q(x) \}=-\infty}^{\infty} \ \sum_{\{ r(l) \}=0}^{\infty} \ \sum_{\{ \rho_1(x),\rho_2(x) \}} \ 
\prod_{x} \left [ \delta_{r(x) + | \eta(x) | - | \tilde{\eta}(x) |, q(x) N} \right ]   \nonumber  \\ 
&\times& \prod_{x} F(\eta(x),\tilde{\eta}(x)) \ \prod_{p_{\rm even }} \ B_p(\rho_i(x))  \ , 
\label{PF_II_puregauge_2d_1}
\end{eqnarray}
where the Boltzmann weight $B_p(\rho_i(x))$ and the function $F(\eta(x),\tilde{\eta}(x))$ are 
\begin{align}
&B_p(\rho_i(x)) =  \sum_{\lambda_1 \vdash r(l_1)}\ldots\sum_{\lambda_{4}\vdash r(l_4)} D_{\lambda_1} (\beta)\ldots D_{\lambda_4} (\beta) 
C^{\rho_1(x)}_{\lambda_1 \ \lambda_4} C^{\rho_1(x+e_1)}_{\bar{\lambda}_1 \ \lambda_2} 
C^{\rho_2(x+e_1+e_2)}_{\lambda_2 \ \lambda_3} C^{\rho_2(x+e_2)}_{\bar{\lambda}_3 \ \lambda_4} , \nonumber  \\ 
\label{Fx_source_2d}
&F(\eta(x),\tilde{\eta}(x)) = \sum_{\sigma} \ C^{\sigma}_{\rho_1(x) \ \eta(x)} \ C^{\sigma+q(x)^N}_{\rho_2(x) \ \tilde{\eta}(x)} \ . 
\end{align}
For the partition function $F(0,0)=\delta_{\rho_2(x),\rho_1(x)+q(x)^N}$ and (\ref{PF_II_puregauge_2d_1}) reduces to 
\begin{eqnarray}
\label{PF_II_puregauge_2d_2}
Z \ = \ \sum_{\{ q(x) \}=-\infty}^{\infty} \ \sum_{\{ r_l \}=0}^{\infty} \ \sum_{\{ \rho(x)\}} \ 
\prod_{x} \delta_{r(x), q(x) N}  \ \prod_{p_{\rm even }} \ B_p(\rho(x))  \ .  
\end{eqnarray}
The Boltzmann weight $B_p(\rho(x))$ coincides with (\ref{Fx_source_2d}) up to replacement $\rho_2(x)\to\rho_1(x)+q(x)^N$. 
The $N$-ality constraint in (\ref{PF_II_puregauge_2d_1}), (\ref{PF_II_puregauge_2d_2}) has the same form as in (\ref{constr_puregauge}). 
Therefore, both for quark--anti-quark sources and for baryon sources, one can use the solution (\ref{dual_var2d}). 
Because the link variables $r(l)$ in (\ref{PF_II_puregauge_2d_1}) take on the non-negative values, the dual variable $k(l)$ 
on the right-hand side of Eq.(\ref{dual_var2d}) is also non-negative and the difference $r(x)-r(x+e_n)$ should be defined modulo $N$. 
The choice of $\eta(l)$ for each case also remains as has been described after Eq.(\ref{dual_var2d}).

\subsection*{\underline{Three-dimensional model}:} 

In three dimensions we divide the lattice in a set of even and odd cubes and 
couple the Schur functions as shown in the right panel of Fig.(\ref{schur_coupl_comb}). 
In this way the summations over original representations $\lambda(l)$ are factorized inside every even cube. 
Moreover, we first couple representations lying in the horizontal plane, then the resulting representations are coupled 
with representation sitting on the vertical links. The final step is to couple the representations obtained with 
representations $\eta(x),\tilde{\eta}(x)$ from correlation functions. This procedure yields for the correlations of 
the general form 
\begin{eqnarray}
\label{PF_II_puregauge_3d_1}
Z(\eta(x),\tilde{\eta}(x)) &=& \sum_{\{ q(x) \}=-\infty}^{\infty} \ \sum_{\{ r(l) \}=0}^{\infty} \ \sum_{\{ \rho_1(x),\rho_2(x) \}} \ 
\prod_{x} \left [ \delta_{r(x) + | \eta(x) | - | \tilde{\eta}(x) |, q(x) N} \right ]   \\ 
&\times& \prod_{x} F(\eta(x),\tilde{\eta}(x)) \ \prod_{c_{\rm even }} \ B_c(\rho_i(x))  \ , 
\end{eqnarray}
where the Boltzmann weight $B_c(\rho_i(x))$ is
\begin{eqnarray}
\label{Bc_puregauge_3d}
&&B_c(\rho_i(x))  =  \sum_{\lambda_1 \vdash r(l_1)} D_{\lambda_1} (\beta) \ldots \sum_{\lambda_{12}\vdash r(l_{12})} D_{\lambda_{12}} (\beta) \ 
\sum_{\sigma_1} \ldots \sum_{\sigma_8}  \\ 
&&\times C^{\sigma_1}_{\lambda_1 \ \lambda_4} \ C^{\sigma_2}_{\lambda_2 \ \bar{\lambda}_1} \ C^{\sigma_3}_{\bar{\lambda}_3 \ \bar{\lambda}_2} \ 
C^{\sigma_4}_{\lambda_3 \ \bar{\lambda}_4} \ 
C^{\sigma_5}_{\bar{\lambda}_5 \ \bar{\lambda}_8} \ C^{\sigma_6}_{\lambda_5 \ \bar{\lambda}_6} \ C^{\sigma_7}_{\lambda_6 \ \lambda_7} \ 
C^{\sigma_8}_{\lambda_8 \ \bar{\lambda}_7} \ 
C^{\rho_1(x)}_{\sigma_1 \ \lambda_9} \ C^{\rho_2(x+e_3)}_{\sigma_5 \ \lambda_9}  \nonumber  \\ 
&&\times C^{\rho_1(x+e_1)}_{\sigma_2 \ \lambda_{10}} \  C^{\rho_2(x+e_1+ e_3)}_{\sigma_6 \ \lambda_{10}} \ 
C^{\rho_1(x+e_1+e_2)}_{\sigma_3 \ \lambda_{11}}  \ C^{\rho_2(x+e_1+e_2+e_3)}_{\sigma_7 \ \lambda_{11}}  \ 
C^{\rho_1(x+e_2)}_{\sigma_4 \ \lambda_{12}}  \ C^{\rho_2(x+e_2+e_3)}_{\sigma_8 \ \lambda_{12}} \ .  \nonumber 
\end{eqnarray}
The function $F(\eta(x),\tilde{\eta}(x))$ is of the form (\ref{Fx_source_2d}).
The specification of this result for two- and $N$-point correlation functions is essentially the same as in the two-dimensional case. 
The solution of $N$-ality constraint is taken as in (\ref{dual_var3d}) with restrictions described after (\ref{PF_II_puregauge_2d_2}). 
The resulting dual model is a three-dimensional model possessing local $Z(N)$ invariance in analogy with (\ref{PFspin_3d_2point}).

\subsection{Strong coupling limit}

In the strong coupling limit, $\beta=0$, the gauge part is absent and only static fermion contribution 
appears in the partition and correlation functions. Essentially, the model is nothing but the one-dimensional lattice QCD. 
The partition function is given by the integral (\ref{Hint_d0})
\begin{eqnarray}
Z \ = \  \sum_{q=-\infty}^{\infty} \ \sum_{\sigma} \  s_{\sigma}(H_+) s_{N^q \sigma}(H_-) \ . 
\label{1dQCD_PF}
\end{eqnarray}
The result of the integration for the correlation function can be easily extracted from (\ref{Hint_d1})  
\begin{eqnarray}
Z(\eta,\tilde{\eta}) \  = \   
\sum_{q=-\infty}^{\infty}  \sum_{\alpha,\beta,\sigma} \ C^{\sigma}_{\eta \ \alpha} C^{\sigma +q^N}_{\tilde{\eta} \ \beta} \ 
s_{\alpha^{\prime}}(H_+) s_{\beta^{\prime}}(H_-) \ . 
\label{1dQCD_corr}
\end{eqnarray}
These two formulas give an exact solution for one-dimensional QCD with arbitrary number of flavours of different masses 
and chemical potentials, both for the staggered and for the Wilson fermions. The detailed investigation of these solutions 
will be presented elsewhere. 

\subsection{Full theory: one-dimension}

One-dimensional model corresponds to two-dimensional QCD. It is important to emphasize that our approach takes into account 
the full Wilson action in this case and the only though essential approximation is that we neglect the fermion propagation in 
one spatial direction. The corresponding one-site integral is given by Eq.(\ref{Hint_d1}). Substituting this into 
(\ref{PF_spindef_II_2}) one gets for the correlation function 
\begin{eqnarray} 
\label{PF_II_full_d1}
&&Z(\eta(x),\tilde{\eta}(x)) = \sum_{\{ q(x) \} =-\infty}^{\infty}  \ \sum_{\{ |\alpha(x)| \}=0}^{c N N_f}  \ 
\sum_{\{ |\beta(x)| \}=0}^{c N N_f} \  
\sum_{\{ r(l) \} = 0}^{\infty} \ \sum_{\{ \rho_1(x), \rho_2(x) \}}    \\
&&\times \prod_x \ \left [ \delta_{r(x) + | \alpha(x) | + | \eta(x) |  - | \beta(x) | - | \tilde{\eta}(x) |, q(x) N} \ 
F(\eta(x),\tilde{\eta}(x)) \right ] \ 
\prod_l \ B_l (\rho_i(x)) \ .  \nonumber   
\end{eqnarray}
The Boltzmann weight and the function $F(\eta(x),\tilde{\eta}(x))$ are found to be 
\begin{eqnarray} 
\label{Bl_II_full_d1}
B_l (\rho_i(x)) =  \sum_{\lambda \vdash r(l)} \sum_{\alpha\vdash |\alpha(x)|} \  \sum_{\beta\vdash |\beta(x+1)|,}  
D_{\lambda}(\beta) \ 
C^{\rho_1(x)}_{\lambda \ \alpha} C^{\rho_2(x+1)}_{\lambda \ \beta} 
s_{\alpha^{\prime}}(H_+) s_{\beta^{\prime}}(H_-)  , 
\end{eqnarray}
\begin{equation}
F(\eta(x),\tilde{\eta}(x)) \ = \ \sum_{\sigma} \ C^{\sigma}_{\rho_1(x) \ \eta(x)} \ C^{\sigma+q(x)^N}_{\rho_2(x) \ \tilde{\eta}(x)} \ . 
\label{Fx_source_full_1d}
\end{equation}
The partition function is easily recovered putting $\eta(x)=\tilde{\eta}(x)=0$.

\subsection{Full theory: two-dimensions}

To get a dual form for two-dimensional theory we use the result of the integration presented in Eqs.(\ref{Hint_d1}), (\ref{Hint_dn}) and 
follow the strategy used in the pure gauge theory. Namely, we first multiply characters from the gauge Boltzmann weight according 
to Fig.(\ref{schur_coupl_comb}), resulting representations are coupled with representations from the fermion weight and, finally 
with representations $\eta(x),\tilde{\eta}(x)$ from the correlation function. After some algebraic manipulations we present 
the expression for the correlation function in the following form 
\begin{eqnarray}
\label{PF_II_full_2d_1}
&&Z(\eta(x),\tilde{\eta}(x)) = \sum_{\{ q(x) \} =-\infty}^{\infty}  \ \sum_{\{ |\alpha(x)| \}=0}^{c N N_f}  \ 
\sum_{\{ |\beta(x)| \}=0}^{c N N_f} \  
\sum_{\{ r(l) \} = 0}^{\infty} \ \sum_{\{ \rho_1(x), \rho_2(x) \}} \\ 
&&\times \prod_x \ \left [ \delta_{r(x) + | \alpha(x) | + | \eta(x) |  - | \beta(x) | - | \tilde{\eta}(x) |, q(x) N} \ 
F(\eta(x),\tilde{\eta}(x)) \right ] \prod_{p_{\rm even}} \ B_p (\rho_i(x)) \ .  \nonumber 
\end{eqnarray}
The function $F(\eta(x),\tilde{\eta}(x))$ is the same as in Eq.(\ref{Fx_source_full_1d}). The Boltzmann weight reads 
\begin{eqnarray}
\label{Bp_full_2d}
&&B_p (\rho_i(x)) =  \sum_{\lambda_1 \vdash r(l_1)}D_{\lambda_1} (\beta) \ldots \sum_{\lambda_{4}\vdash r(l_4)} D_{\lambda_4} (\beta) 
\sum_{\sigma_1 } \ldots \sum_{\sigma_4} 
C^{\sigma_1}_{\lambda_1 \ \lambda_4} C^{\sigma_2}_{\bar{\lambda}_1 \ \lambda_2}C^{\sigma_3}_{\lambda_2 \ \lambda_3} 
C^{\sigma_4}_{\bar{\lambda}_3 \ \lambda_4}  \nonumber \\
&&  \sum_{\alpha_1\vdash |\alpha(x)|}  s_{\alpha_1^{\prime}}(H_+) \sum_{\alpha_2\vdash |\alpha(x+e_1)|} \ s_{\alpha_2^{\prime}}(H_+) 
\sum_{\beta_1\vdash |\beta(x+e_2)|}  s_{\beta_1^{\prime}}(H_-) \ \sum_{\beta_2\vdash |\beta(x+e_1+e_2)|}  s_{\beta_2^{\prime}}(H_-) \nonumber  \\  
&&C^{\rho_1(x)}_{\sigma_1 \ \alpha_1} \ C^{\rho_1(x+e_1)}_{\sigma_2 \ \alpha_2} \ 
 C^{\rho_2(x+e_1+e_2)}_{\sigma_3 \ \beta_1} \ C^{\rho_2(x+e_2)}_{\sigma_4 \ \beta_2}  \ . 
\end{eqnarray}

\subsection{Full theory: three-dimensions}

In three-dimensional theory we use the same strategy as above. The original link representations are coupled as shown  
in Fig.(\ref{schur_coupl_comb}). With the help of Eqs.(\ref{Hint_d1}), (\ref{Hint_dn}) and using the same notations 
we obtain after long algebra the following representation for the correlation function 
\begin{eqnarray}
\label{PF_II_full_3d_1}
&&Z(\eta(x),\tilde{\eta}(x)) = \sum_{\{ q(x) \} =-\infty}^{\infty}  \ \sum_{\{ |\alpha(x)| \}=0}^{c N N_f}  \ 
\sum_{\{ |\beta(x)| \}=0}^{c N N_f} \  
\sum_{\{ r(l) \} = 0}^{\infty} \ \sum_{\{ \rho_1(x), \rho_2(x) \}} \\ 
&&\times \prod_x \ \left [ \delta_{r(x) + | \alpha(x) | + | \eta(x) |  - | \beta(x) | - | \tilde{\eta}(x) |, q(x) N} \ 
F(\eta(x),\tilde{\eta}(x)) \right ] \prod_{c_{\rm even}} \ B_c (\rho_i(x)) \ .   \nonumber  
\end{eqnarray}
The function $F(\eta(x),\tilde{\eta}(x))$ is the same as in Eq.(\ref{Fx_source_full_1d}). The Boltzmann weight reads 
\begin{eqnarray}
&&B_c(\rho_i(x)) =  \sum_{\lambda_1 \vdash r(l_1)}D_{\lambda_1} (\beta) \ldots 
\sum_{\lambda_{12}\vdash r(l_{12})} D_{\lambda_{12}} (\beta) \ \sum_{\sigma_1,\gamma_1}  \ldots  \sum_{\sigma_8,\gamma_8}  \nonumber  \\
&&C^{\sigma_1}_{\lambda_1 \ \lambda_4} \ C^{\sigma_2}_{\lambda_2 \ \bar{\lambda}_1} \ C^{\sigma_3}_{\bar{\lambda}_3 \ \bar{\lambda}_2} \ 
C^{\sigma_4}_{\lambda_3 \ \bar{\lambda}_4} \ 
C^{\sigma_5}_{\bar{\lambda}_5 \ \bar{\lambda}_8} \ C^{\sigma_6}_{\lambda_5 \ \bar{\lambda}_6} \ C^{\sigma_7}_{\lambda_6 \ \lambda_7} \ 
C^{\sigma_8}_{\lambda_8 \ \bar{\lambda}_7} \ \nonumber  \\
\label{Bc_full_3d}
&& C^{\gamma_1}_{\sigma_1 \ \lambda_9 } \ C^{\gamma_5}_{\sigma_5 \ \lambda_9 }  \
C^{\gamma_2}_{\sigma_2 \ \lambda_{10} } \ C^{\gamma_6}_{\sigma_6 \ \lambda_{10}} \
C^{\gamma_3}_{\sigma_3 \ \lambda_{11}}  \ C^{\gamma_7 }_{\sigma_7 \ \lambda_{11}} \ 
C^{\gamma_4 }_{\sigma_4 \ \lambda_{12}} \ C^{\gamma_8 }_{\sigma_8 \ \lambda_{12}}   \\
&&\prod_{i=1}^4 \left [ \sum_{\alpha_i\vdash |\alpha(x_i)|}  s_{\alpha_i^{\prime}}(H_+) \ 
C^{\rho_1(x_i)}_{ \gamma_i \ \alpha_i} \right ] \  
\prod_{i=5}^8 \left [ \sum_{\beta_i\vdash |\beta(x_i)|}  s_{\beta_i^{\prime}}(H_-) \ 
C^{\rho_2(x_i)}_{\gamma_i \ \beta_i}  \right ]  \  . \nonumber  
\end{eqnarray}
Notation for sites of a cube are: $x_1=x,x_2=x+e_1,x_3=x+e_1+e_2,x_4=x+e_2,x_5=x+e_3,x_6=x+e_1+e_3,x_7=x+e_1+e_2+e_3,x_8=x+e_2+e_3$.  

As an important example, let us write down an explicit formula for the partition function with one flavour of 
the staggered fermions. In this case one has $\alpha=1^k$, $\beta=1^m$, $0\leq k,m \leq N$ and $s_{\alpha^{\prime}}(H_+)=h_+^k$, 
$s_{\beta^{\prime}}(H_-)=h_-^m$, therefore the Eq.(\ref{PF_II_full_3d_1}) takes the form for $\eta(x)=\tilde{\eta}(x)=0$ 
\begin{eqnarray}
\label{PF_II_full_3d_stag}
&&Z(\eta(x),\tilde{\eta}(x)) = \sum_{\{ q(x) \} =-\infty}^{\infty}  \  \sum_{\{ r(l) \} = 0}^{\infty} \ 
\sum_{\{ \rho(x) \}} \ \sum_{\{k(x),m(x) \}=0}^{N} \prod_x h_+^{k(x)} h_-^{m(x)}  \\ 
&&\times \prod_x \ \left [ \delta_{r(x) + k(x) - m(x), q(x) N} \ \right ] 
\prod_{c_{\rm even}} \ B_c (\rho(x)) \ , \nonumber 
\end{eqnarray}
In this case the Boltzmann weight (\ref{Bc_full_3d}) becomes 
\begin{eqnarray}
&&B_c(\rho_i(x)) =  \sum_{\lambda_1 \vdash r(l_1)}D_{\lambda_1} (\beta) \ldots 
\sum_{\lambda_{12}\vdash r(l_{12})} D_{\lambda_{12}} (\beta) \ \sum_{\sigma_1,\gamma_1}  \ldots  \sum_{\sigma_8,\gamma_8}  \nonumber  \\
&&\times C^{\sigma_1}_{\lambda_1 \ \lambda_4} \ C^{\sigma_2}_{\lambda_2 \ \bar{\lambda}_1} \ C^{\sigma_3}_{\bar{\lambda}_3 \ \bar{\lambda}_2} \ 
C^{\sigma_4}_{\lambda_3 \ \bar{\lambda}_4} \ 
C^{\sigma_5}_{\bar{\lambda}_5 \ \bar{\lambda}_8} \ C^{\sigma_6}_{\lambda_5 \ \bar{\lambda}_6} \ C^{\sigma_7}_{\lambda_6 \ \lambda_7} \ 
C^{\sigma_8}_{\lambda_8 \ \bar{\lambda}_7} \   \nonumber  \\
\label{Bc_full_3d_stag}
&& C^{\gamma_1}_{\sigma_1 \ \lambda_9 } \ C^{\gamma_5}_{\sigma_5 \ \lambda_9 }  \
C^{\gamma_2}_{\sigma_2 \ \lambda_{10} } \ C^{\gamma_6}_{\sigma_6 \ \lambda_{10}} \
C^{\gamma_3}_{\sigma_3 \ \lambda_{11}}  \ C^{\gamma_7 }_{\sigma_7 \ \lambda_{11}} \ 
C^{\gamma_4 }_{\sigma_4 \ \lambda_{12}} \ C^{\gamma_8 }_{\sigma_8 \ \lambda_{12}}   \\
&&\prod_{i=1}^4 C^{\rho(x_i)}_{ \gamma_i \ 1^{k(x_i)}}  \  
\prod_{i=5}^8 C^{\rho(x_i)+q(x_i)^N}_{\gamma_i \ 1^{m(x_i)}} \  .   \nonumber 
\end{eqnarray}

Let us add some remarks on resulting dual formulations: 

\begin{itemize}
\item 
All factors entering the dual Boltzmann weights obtained in this Section are positive including 
the Schur functions. Hence, this representation is suitable for the Monte-Carlo simulations at non-vanishing 
chemical potentials, at least in principle. Possible approaches to such simulations are discussed 
in the end of this section. 

\item
One technical remark concerns the function $F(\eta(x),\tilde{\eta}(x))$ which appears in the presence of external 
sources for the correlation function. From its explicit expresssion, Eqs.(\ref{Fx_source_2d}) and (\ref{Fx_source_full_1d}), 
it follows that the product over lattice sites can be rearranged in a way that allows to include all Littlewood-Richardson 
coefficients from $F(\eta(x),\tilde{\eta}(x))$ into the Boltzmann weights. In this way the variables $\rho_i(x)$ become 
internal summation variables in each even plaquette (cube). Instead, variables $\sigma(x)$ can be made dynamical variables 
of the dual theory. 

\item 
We have not presented explicit expressions for the thermodynamical quantities. They can be easily obtained 
by taking the corresponding derivatives with respect to $\beta$, $h_{\pm}^f$ 
or $\mu_f$. As in the region of the strong temporal coupling, this simply amounts to a local shift 
in a corresponding summation variable and can be presented as an expectation value 
calculated over the dual partition function. 

\item 
An explicit form of the group integral $H_{N,N_f}^d$ for one flavour of the staggered fermions 
is presented in Appendix, Section (\ref{stag_1fl}). 

\item 
The dual forms of $U(N)$ and $Z(N)$ models can be obtained following the lines described in the previous section. 
It follows from the $N$-ality constraint in Eq.(\ref{PF_II_full_3d_stag}) that the partition function and invariant observables 
do not depend on the chemical potential for $U(N)$ models with one fermion flavour.
In $SU(N)$ and $Z(N)$ models the dependence appears in the form $e^{N q \mu}$, as is expected on the general grounds.

\end{itemize}

Finally, we would like to discuss shortly some approaches to Monte-Carlo simulations of the dual formulations 
and related problems. 
Since the Boltzmann weights for the dual model obtained in this paper
are nonnegative, they can be used for direct Monte-Carlo simulations.
Dual weights based on the expansion of the group integrals into the Littlewood-Richardson coefficients 
might look quite complicated at first sight. Let us remark, however that if a positive dual weight 
at finite density exists, for the full theory it will certainly be much more complicated. Therefore, 
it is desirable to have a working algorithm for this complicated but still simplified dual theory.
Here we give our thoughts on the way Monte-Carlo simulations can be performed.
We will address explicitly the case of the partition functions
(\ref{PF_II_full_d1}), (\ref{PF_II_full_2d_1}), (\ref{PF_II_full_3d_1})
but the approaches described can also be applied to other cases.

The partition function as it is written includes summation over many
sets of variables. While a direct Monte-Carlo simulation is possible,
the convergence of the averages in this case can be slow. This can be
overcome by dividing the variables into two groups -- dynamical variables
which are sampled in the Monte-Carlo simulation, and the variables over
which the summation is done explicitly. For $N=3$ such summation can be
done by using the explicit values for the Littlewood-Richardson
coefficients \cite{su3_multiplicity}, \cite{LR_coeff}.
If we take this approach, we can, for example, leave only
$|\alpha(x)|$, $|\beta(x)|$ and $\rho(x)$ variables as dynamical.

Another problem is that, while each configuration has nonnegative weight,
many configurations formally allowed in the summation will have zero weight
due to Littlewood-Richardson coefficients inside them becoming zero.
Note that the $N$-ality condition explicitly written in the partition
functions is a necessary but not sufficient condition for the corresponding
Littlewood-Richardson coefficients to be nonzero.
This problem reduces the acceptance and convergence rate for any simple
Metropolis-like update scheme and, on the more fundamental level,
raises the question of ergodicity of the update process -- one has to be
sure that the whole acceptable configuration space can be probed.

For fixed small $N$ values the full set of conditions
for the Littlewood-Richardson coefficient to be nonzero can be written
in the form of inequalities and one can either try to explicitly resolve
them, or, at least, to build the update process respecting
these inequalities.
Another approach is to use the worm update algorithm\cite{worm},
developed just for resolving such problems.
In our case the worm has to propagate on an auxiliary lattice, that has
the Littlewood-Richardson coefficients of the partition function as
vertices, which are connected by a link if they share a common partition.
Like for the Metropolis-like update, an explicit summation over part of
partition variables can be done to reduce the phase space of the system --
here it would amount to dividing the auxiliary lattice into blocks that
are connected only by the links corresponding to the dynamical variables and
treating each block as a site of a new lattice, while preserving the
probabilities
of the worm leaving the block through a given link.

If one wants to calculate the correlation functions, the set of
acceptable configurations for the partition functions with sources
and the one without sources will become different,
requiring either rewriting  the correlation function in terms
of the valid configurations for the partition function without sources,
or, if one uses the worm update, sampling the correlation functions
 using the worm algorithm in a way similar to the one described in
\cite{worm-corr}.

\section{Discussion}

In this paper we have presented calculations of the dual representations for several 
Polyakov loop models. All these models have been derived in the strong coupling limit for the spatial 
coupling of the Wilson action. Contribution of the fermions is taken into account via the static 
determinant for an arbitrary number of the staggered and Wilson fermions of different masses 
and chemical potentials. Our results are valid in any spatial dimension and for all relevant 
groups, {\it i.e.} for $SU(N)$, $U(N)$ and $Z(N)$. The main motivation is the construction of 
a positive Boltzmann weight in the presence of the baryon chemical potential suitable for 
numerical simulations. Some versions of representations from Sect.~3 have been derived before \cite{spin_flux1}. 
These formulations have been used for numerical computations of various local observables in \cite{spin_flux2}, 
\cite{philipsen12}. We have already applied our formulation (\ref{PFspin_2d_2point}) for studying 
two- and three-point correlation functions in two dimensional $SU(3)$ spin models \cite{2d_baryon}. 
In \cite{3d_su3} we have used the representation (\ref{PF_statdet_3}) to simulate the three-dimensional 
model with one flavour of the staggered fermions. In addition to local observables we have computed 
many two-point correlation functions in the presence of baryon chemical potential. 

Let us briefly discuss other applications of the dual formulations. 
\begin{enumerate}
\item 
One-dimensional model (\ref{PF_II_full_d1}) with one flavour of the staggered fermions can be studied 
by the transfer-matrix method. Such study reveals the existence of an oscillating (or the so-called liquid) 
phase in some regions of the $(h_+,h_-)$-plane \cite{Ogilvie16}. This means the correlation function 
of the Polyakov loops while decaying exponentially is modulated by a periodic function. In other words, 
the mass spectrum of the theory becomes complex. The transfer matrix approach reveals the similar behaviour 
in one-dimensional $Z(N)$ spin model in the external complex field \cite{Ogilvie10}, \cite{oscillating_phase}. 
Monte-Carlo simulations of the same three-dimensional model also show the presence of such phase \cite{oscillating_phase}. 
Detecting the liquid phase with the existing simulation methods at non-zero baryon chemical potential 
seems an extremely difficult problem. The formulations given in Sect.~3 might help to clarify if the oscillating phase 
exists in the three-dimensional $SU(3)$ LGT, at least in the region of validity of the Polyakov loop models used here.

\item 
It turned out that the dual formulations of Sect.~3 are well suited for the studies of the models in the large $N$ limit. 
We have accomplished such studies and arrived at quite unexpected results: the large $N$ 't Hooft limits 
of $U(N)$ and $SU(N)$ models are different in the presence of the chemical potentials. 
These results will be published elsewhere. 

\item 
The partition function in Eq.(\ref{PF_spindef_II_2}) can be written at zero sources as 
\begin{equation}
\label{tensor_repr_PF}
Z \ = \ {\rm Tr} \ \prod_{x} \ T_{\lambda(l_1)\cdots\lambda(l_{2d})}(\beta, h_+^f, h_-^f) \ , 
\end{equation}
where $2d$ links $l_1,\ldots,l_{2d}$ are attached to a site $x$ and the tensor $T$ reads 
\begin{equation}
T_{\lambda(l_1)\cdots\lambda(l_{2d})}(\beta, h_+^f, h_-^f) \ = \ \prod_{\nu=1}^{2d}
\left [ D_{\lambda(l)}(\beta) \right ]^{\frac{1}{2}} \ H_{N,N_f}^d(g(x),f(x);m_f,\mu_f) \ . 
\label{tensor_T}
\end{equation}
The rank of the tensor is $2d N$. The trace can be done by properly contracting the indices 
labelling group representations. Such a formulation enables one to use the tensor renormalization 
group methods to study the theory at finite density and is certainly worth a separate investigation. 

\item 
It would be interesting to investigate analytically solutions (\ref{1dQCD_PF}) and (\ref{1dQCD_corr}) 
of one-dimensional QCD for small values of $N=1,2,3$ and in the large $N,N_f$  limits. This can be presumably 
done even in the simultaneous presence of baryon, isospin and strange chemical potentials. 

\end{enumerate}

Finishing this paper we would like to address the question of how one could systematically improve 
the dual formulations of Sect.~4? In the abelian case the dual construction can be extended to 
the full Wilson action. The underlying reason for this is that the exact and positive dual Boltzmann weight 
is known for all $Z(N)$ and $U(1)$ pure gauge LGT. Adding fermions in the form of the static determinant 
with any number of flavours does not destroy this property. Details of this formulation will be reported elsewhere. 
Much more difficult is the case of non-abelian models and the inclusion of the corrections to static determinant. 
In these cases one expects that the effective Polyakov loop model becomes non-local. For example, 
one such model describing non-local interaction between Polyakov loops has been derived in \cite{greensite16} 
(and Refs. therein) via the relative weight method
\begin{equation}
S \ = \ \sum_{x,y} \ {\rm Re} \ {\rm Tr} U(x)  K(x-y) {\rm Tr} U^{\dagger}(y) + 
\sum_x \ln B_q(m,\mu) \ . 
\label{nonlocal_PL}
\end{equation}
The action of the model involves only fundamental characters, therefore the dual representations 
for this and similar models can be calculated by using the integration methods of Sect.~3. 
Clearly, if the kernel $K(x-y)$ is positive for all distances considered the dual weight will be also positive 
but highly non-local. In general, the full effective action will contain all irreducible representations. 
More general effective action can be written in the form 
\begin{equation}
e^S \ = \ \prod_{x,y} \ \left [ \sum_{\lambda,\gamma} \  K_{\lambda,\gamma}(x-y; \beta, m_f, \mu_f) \ 
s_{\lambda}(U(x)) s_{\gamma}(U^{\dagger}(y)) \right ] \ . 
\label{gen_nonlocal_PL}
\end{equation}
Even in this case the dual theory could be calculated with the help of integration methods of Sect.~4. 
The real challenge is to determine coefficients $K_{\lambda,\gamma}$. One strategy is to expand the Wilson action at large 
spatial coupling and expand the fermion determinant around static contribution in powers of a lattice anisotropy. 
This will be the subject of future investigations.

\section*{Appendix}

Here we evaluate the group integrals encountered in the course of calculations 
of the dual representations. These are integrals of the following types: 
\begin{equation} 
Q_N(r,s) = \int_G dU \ ({\rm Tr} U)^r \ ({\rm Tr} U^{\dagger})^s  \ , 
\label{Q_def}
\end{equation} 
\begin{eqnarray}
G_N^d(\lambda_i,\gamma_i )  =  \int_G dU \ \prod_{i=1}^d \  s_{\lambda_i}(U) \ s_{\gamma_i}(U^{\dagger}) \ , 
\label{Gint_def}
\end{eqnarray}
\begin{equation} 
R_{N,N_f}(r,s;m_f,\mu_f) = \int_G dU \ ({\rm Tr} U)^r \ ({\rm Tr} U^{\dagger})^s  \ 
\prod_{f=1}^{N_f} \ B_q(m_f,\mu_f) \ , 
\label{Rint_def}
\end{equation}
\begin{eqnarray}
H_{N,N_f}^d(\lambda_i,\gamma_i;m_f,\mu_f)  =  \int_G dU \ \prod_{i=1}^d \  s_{\lambda_i}(U) \ s_{\gamma_i}(U^{\dagger}) \ 
\prod_{f=1}^{N_f} \ B_q(m_f,\mu_f) \ .
\label{Hint_def}
\end{eqnarray}
The first two types appear in the pure gauge theory while the next two are encountered in the theory with 
the fermions. The fermion weights $B_q(m_f,\mu_f)$ are given by Eqs. (\ref{Zf_stag}) and (\ref{Zf_wilson}). 
These integrals can be calculated by the method of the Weingarten functions and/or by expanding 
the product of the Schur functions in the integrand into series over the Littlewood-Richardson coefficients.

\subsection*{A1: Definitions, notations and expansion formulas}

First, we introduce some notations and definitions. 
Let $\lambda=(\lambda_1,\lambda_2,\cdots,\lambda_N)$ be a partition $\lambda\vdash\ r$, 
$\lambda_1\geq\lambda_2\geq \cdots \geq \lambda_N\geq 0$ and $\sum_{i=1}^{l(\lambda)} \lambda_i \equiv |\lambda|= r$, 
where $l(\lambda)$ is the length of the partition $\lambda$. 
As a shorthand we will sometimes use a notation $\lambda = a^b$ to denote a partition consisting of $b$ parts equal to $a$
({\it i.e.}, $\lambda_i = a, 1\leqslant i \leqslant b$), and use $\lambda + \mu$ to signify elementwise addition of two partitions 
and $\lambda \mu$ to signify a union of parts of two partitions.
$\chi_{\lambda}(\sigma)$ denotes a character of $\sigma \in S_r$ 
in representation $\lambda$. $d(\lambda)=\chi_{\lambda}(1)$ is the dimension of the representation $\lambda$. 
The Schur function $s_{\lambda}(U)=s_{\lambda}(u_1,\cdots,u_N)$ is a character of the unitary group $G$ and  
$u_i$ - the eigenvalues of the matrix $U\in G$. $s_{\lambda}(I)$ is 
the  dimension of the irreducible representation $\lambda$ of $G$. One has 
\begin{equation}
d(\lambda) \ = \ r! \ 
\frac{\prod_{1\leq i<j \leq l(\lambda)}(\lambda_i-\lambda_j+j-i)}{\prod_{i=1}^{l(\lambda)}(\lambda_i+l(\lambda)-i)!} \ , 
\label{dim_l}
\end{equation}
\begin{equation}
s_{\lambda}(I) \ = \ \frac{\prod_{1\leq i<j \leq N}(\lambda_i-\lambda_j+j-i)}{\prod_{i=1}^N( N-i)!} \ . 
\label{G_dim}
\end{equation} 
The representation dual to $\lambda$ will be denoted by $\lambda^{\prime}$. The dual representation is defined by exchanging 
raws and columns in the corresponding Young diagram, {\it i.e.} $\lambda^{\prime}_i=\sum_j \mathbf{1}_{\lambda_j\geq i}$. 
One has the following identity between the Schur function and its conjugate for $U(N)$ group 
\begin{equation}
s_{\lambda}(U) \ \equiv \ s_{\lambda_1,\cdots ,\lambda_N}(U) \ = \ s^*_{\bar{\lambda}}(U) \ , \ 
\bar{\lambda} \ = \ (-\lambda_N, -\lambda_{N-1},\cdots , -\lambda_1 ) \ . 
\label{schur_conj_un}
\end{equation}
The similar identity for $SU(N)$ group reads 
\begin{equation}
s_{\lambda}(U) \ \equiv \ s_{\lambda_1,\cdots ,\lambda_{N-1}}(U) \ = \ s^*_{\bar{\lambda}}(U) \ , \ 
\bar{\lambda} \ = \ (\lambda_1, \lambda_1-\lambda_{N-1},\cdots , \lambda_1-\lambda_2 ) \ . 
\label{schur_conj_sun}
\end{equation}

Given the complete symmetric functions $h_k$ 
and the elementary symmetric functions $e_k$ in $m$ variables $u_1,\cdots ,u_m$ 
\begin{eqnarray}
h_k \ = \ \sum_{1\leq n_1\leq\cdots\leq n_k \leq m} 
\ u_{n_1} \cdots u_{n_k} \ , \ 
e_k \ = \ \sum_{1\leq n_1 < \cdots < n_k \leq m} \ u_{n_1} \cdots u_{n_k} \ , 
\label{ef_hf}
\end{eqnarray} 
the Schur functions can be computed with the help of identities 
\begin{eqnarray} 
s_{\lambda}(U) \ = \ s_{(\lambda_1,\cdots,\lambda_m )}(u_1,\cdots,u_m) \ = \ 
\mbox{det} 
\left ( h_{\lambda_i-i+j}  \right )_{1\leq i,j \leq m} \ , \nonumber  \\ 
s_{\lambda}(U) \ = \ s_{(\lambda_1,\cdots,\lambda_m )}(u_1,\cdots,u_m) \ = \ 
\mbox{det} \left ( e_{\lambda^{\prime}_i-i+j}  \right )_{1\leq i,j \leq m} \ , 
\label{ef_hf_schur}
\end{eqnarray}
where $\lambda^{\prime}$ is a partition dual to
$\lambda$ and the following rule is understood 
\begin{equation}
s_{(\lambda_1,\cdots,\lambda_n)}(u_1,\cdots,u_{n-1},0) \ = \   
\begin{cases}
0, \ {\rm {if}} \ \lambda_n \ne 0  \ \ , \\ 
s_{(\lambda_1,\cdots,\lambda_{n-1})}(u_1,\cdots,u_{n-1}), 
\ {\rm {if}} \ \lambda_n=0 \ . 
\end{cases}
\label{schur_reduction}
\end{equation} 
Two other useful expressions for the Schur function read 
\begin{equation} 
s_{\lambda}(U) = \frac{\mbox{det} \left(u_i^{\lambda_j + n - j}\right )_{1\leq i,j \leq N}}
{\mbox{det} \left(u_i^{n - j}\right )_{1\leq i,j \leq N}}
= \sum_{\tau_1,...,\tau_s; \sum i\tau_i=s} 
\ \chi_{\lambda}(\tau) 
\prod_{j=1}^s \ \frac{1}{\tau_j! j^{\tau_j}}  
\  \left [  {\rm Tr} (U)^j \right ]^{\tau_j} 
\label{schur_repr1}
\end{equation}
with $\tau$ being a permutation such that the number of cylcles of 
length $j$ in $\tau$ is $\tau_j$. 

The Weingarten function is used to evaluate the polynomial integrals over unitary groups. 
In the case of $G=SU(N)$ this function is defined as 
\begin{equation} 
Wg^{N,q}(\sigma) = \frac{1}{r!(r+Nq)!} \ \sum_{\lambda\vdash\ r} \ 
\frac{d(\lambda)d(\lambda + q^N)}{s_{\lambda}(1^N)} \ \chi_{\lambda}(\sigma) \ , 
\label{Wgn_def}
\end{equation} 
where $\lambda+q^N=(\lambda_1+q,\cdots,\lambda_N+q)$ and the sum in (\ref{Wgn_def}) is taken over all 
$\lambda$ such that $l(\lambda)\leq N$. For $U(N)$ group one has to put $q=0$.

The Littlewood-Richardson coefficients $C^{\nu}_{\lambda \ \gamma}$ can be defined as coefficients appearing 
in the expansion of the product of two Schur functions 
\begin{eqnarray}
\label{LR_expansion}
s_{\lambda}(U) s_{\gamma}(U) \ = \ \sum_{\nu} \ C^{\nu}_{\lambda \ \gamma} \ s_{\nu}(U) \ ,   \\ 
s_{\lambda}(U) s^*_{\gamma}(U) \ = \ \sum_{\nu} \ C^{\nu}_{\lambda \ \bar{\gamma}} \ s_{\nu}(U) \ = \ 
\sum_{\nu} \ C^{\nu}_{\bar{\lambda} \ \gamma} \ s^*_{\nu}(U) \ .  \nonumber 
\end{eqnarray}
From the orthogonality of the Schur functions one gets 
\begin{equation}
C^{\nu}_{\lambda \ \gamma} \ = \int_G \ dU \ \ s_{\lambda}(U) s_{\gamma}(U) s^*_{\nu}(U) \ . 
\label{LR_int}
\end{equation} 
$C^{\nu}_{\lambda \ \gamma}$ are positive integers for unitary groups $U(N)$ and $SU(N)$ satisfying 
certain conditions. One such important condition on $C^{\nu}_{\lambda \ \gamma}$ to be non-zero arises from 
the integration over $U(1)$ subgroup if $G=U(N)$, while in $SU(N)$ case it follows from the summation over $Z(N)$ subgroup.
Let $U=Z$ be a center element of $G$. Then 
\begin{eqnarray}
s_{\lambda}(Z) \ = \ z^r d(\lambda) \ , \ r = |\lambda| \ . 
\label{center_schur}
\end{eqnarray}
As follows from (\ref{LR_int}) the necessary condition for $C^{\nu}_{\lambda \ \gamma}$ to be 
non-vanishing is 
\begin{eqnarray} 
|\lambda| + |\gamma| - |\nu| \ = \ 
\begin{cases}
 0 \ , \ U(N) \ , \\ 
 N q \ , \ SU(N) \ . 
\end{cases}
\label{LR_cond}
\end{eqnarray}
We refer to these conditions as to $U(1)$ and $N$-ality constraints, respectively. 
More information on the Littlewood-Richardson coefficients as well as their closed forms 
for $N=2,3,4$ can be found in \cite{LR_coeff}. 

Finally, we need the following formulas to treat the models with static fermion determinant.
The first ones are the Cauchy identity and its dual 
\begin{equation}
\prod_{k=1}^N \ \prod_{i=1}^{L} \ \left ( 1- x_i y_k \right )^{-1} \ = \ 
\sum_{\lambda} \  s_{\lambda}(Y) \ s_{\lambda}(X) \ , 
\label{det_schurexp_direct}
\end{equation}
\begin{equation}
\prod_{k=1}^N \ \prod_{i=1}^{L} \ \left ( 1+ x_i y_k \right ) \ = \ 
\sum_{\lambda} \  s_{\lambda}(Y) \ s_{\lambda^{\prime}}(X) \ , 
\label{det_schurexp}
\end{equation}
where $X=(x_1,\cdots,x_L)$, $Y=(y_1,\cdots,y_N)$. The summation over $\lambda$ runs over all partitions 
of $NL$ such that $l(\lambda)\leq N$ and $l(\lambda^{\prime})\leq L$. 
The second one is an expansion of powers of the fundamental character into series over the Schur functions  
\begin{equation}
({\rm Tr} U)^r = (u_1+u_2+\dots+u_N)^r = \sum_{\lambda \vdash r} d(\lambda)s_{\lambda}(U) \ . 
\label{trace_power}
\end{equation}
With the help of Eq.(\ref{det_schurexp}) the fermion contribution given in (\ref{Zf_stag}) and in (\ref{Zf_wilson}) 
for the staggered and the Wilson fermions, respectively, is presented in the form 
\begin{equation}
\prod_{f=1}^{N_f} \ B_q(m_f,\mu_f) \ = \ \sum_{\alpha,\beta} \ s_{\alpha}(U) s_{\beta}(U^{\dagger}) \ 
s_{\alpha^{\prime}}(H_+) s_{\beta^{\prime}}(H_-) \ .
\label{fermdet_exp}
\end{equation} 
For the staggered fermions one has $H_{\pm}=(h_{\pm}^1,\cdots,h_{\pm}^{N_f})$.
The summation over $\alpha$ and $\beta$ is taken over all partitions such 
that $l(\lambda)\leq N$, $l(\beta)\leq N$ and $l(\lambda^{\prime})\leq N_f$, $l(\beta^{\prime})\leq N_f$.
For the Wilson fermions one has $H_{\pm}=(h_{\pm}^1,\cdots,h_{\pm}^{N_f},h_{\pm}^1,\cdots,h_{\pm}^{N_f})$.
The summation over $\alpha$ and $\beta$ is taken over all partitions such 
that $l(\lambda)\leq N$, $l(\beta)\leq N$ and $l(\lambda^{\prime})\leq 2 N_f$, $l(\beta^{\prime})\leq 2 N_f$.
Constants $h_{\pm}^f$ are defined in (\ref{hpm_stag}) and (\ref{hpm_wilson}) for the staggered and the Wilson fermions, 
correspondingly.

\subsection*{A2: Group integrals}

The Schur functions realize representations of $U(N)$ group. Therefore, all integrals 
(\ref{Q_def})-(\ref{Hint_def}) are evaluated over the $U(N)$ Haar measure. If $G=SU(N)$ one should 
introduce an additional constraint into the measure 
\begin{equation}
 {\rm det} \ U \ = \ \prod_{i=1}^N \ u_i \ = \ 1 \ . 
\label{sun_det}
\end{equation}
This constraint can be implemented into the group integrals 
by multiplying the integrand with the delta-function $\sum_{q=-\infty}^{\infty} \left({\rm det} \ U\right)^q$. 
Taking into account that 
\begin{equation}
\left ( \prod_{i=1}^N \ u_i \right )^q \ = \ s_{q,\ldots,q}(U)
\label{det_shurfunc}
\end{equation}
one can easily prove with the help of (\ref{schur_repr1}) that 
\begin{equation}
	\left({\rm det} \ U\right)^q s_\lambda(U) \ = \ s_{\lambda+q^N}(U) \ , \ q > 0 \ . 
\label{schur_shift}
\end{equation}
If $q<0$ one should replace the eigenvalues $u_i$ by $u_i^{*}$. 
Here and further we use the short-hand notation $\lambda+q^N=(\lambda_1+q,\cdots,\lambda_N+q)$. 
The $SU(N)$ constraint is enforced in the formulas below by summation over $q$. 
Furthermore, we shall present results only for the $SU(N)$ group. The $U(N)$ case is easily recovered by 
omitting all sums over $q$ and taking $q=0$ in all formulas below. 
More relevant information on the group integration and similar integrals can be found 
in Refs. \cite{lat_rev}, \cite{carlsson08}, \cite{weingarten_sun}. 

\subsection*{\underline{$Q_N(r,s)$}:}

To evaluate $Q_N(r,s)$ given by Eq.(\ref{Q_def})
we expand the traces in the integrand as sums over diagonal elements
\begin{equation}
({\rm Tr} U)^r \ = \ \sum_{i_1=1}^N \ \cdots   \sum_{i_r=1}^N \ 
\prod_{k=1}^r \ U_{i_ki_k}  \ .
\label{race_exp}
\end{equation}
For the integral (\ref{Q_def}) this leads to
\begin{eqnarray}
Q_N(r,s) \ = \  \sum_{i_1i_2\cdots i_r=1}^N \ \sum_{j_1j_2\cdots j_s=1}^N \ 
\int_G dU \ \prod_{k=1}^r \ U_{i_ki_k} \prod_{m=1}^s U_{j_mj_m}^*   \ . 
\label{trace_intg1}
\end{eqnarray}
The last integral can be calculated with the help of the Weingarten function. 
The details of the derivation can be found in \cite{un_dual,weingarten_sun}. 
Performing summation over group indices one gets
\begin{equation} 
Q_N(r,s) \ = \  
\sum_{q=-\infty}^{\infty} \delta_{r-s,q N} \  
\sum_{\lambda \vdash {\rm min}(r,s)} \ d(\lambda) \ d(\lambda + |q|^N) \ . 
\label{QSUN}
\end{equation}
Another way to compute (\ref{Q_def}) is to use the expansion (\ref{trace_power}). 
Then, the result (\ref{QSUN}) follows from the orthogonality of the Schur functions. 

\subsection*{\underline{$G_N^d(\lambda_i,\gamma_i )$}:}

Integral in (\ref{Gint_def}) is trivial in one-dimensional case, $d=1$, due to orthogonality of the Schur functions 
\begin{equation}
G_N^1(\lambda,\gamma ) \ = \ 
\sum_{q=-\infty}^{\infty} \delta_{\lambda,\gamma + q^N}  \ . 
\label{GN1}
\end{equation}
For $d=2,3$ these integrals can be computed with the help of Eqs.(\ref{LR_expansion})-(\ref{LR_int}). 
Depending on the order of the multiplication of the Schur functions in the integrand the final result 
can be presented in several different but equivalent forms. For example, for $d=2$ one has 
\begin{eqnarray}
G_N^2(\lambda_1,\lambda_2,\gamma_1,\gamma_2 ) &=& 
\sum_q \sum_{\nu} \ C^{\nu}_{\lambda_1 \ \lambda_2} \ C^{\nu + q^N}_{\gamma_1 \ \gamma_2} \nonumber    \\ 
= \sum_q \sum_{\nu} \ C^{\nu}_{\lambda_1 \ \bar{\gamma}_1} \ C^{\nu + q^N}_{\gamma_2 \ \bar{\lambda}_2} &=&   
\sum_q \sum_{\nu} \ C^{\nu}_{\lambda_1 \ \bar{\gamma}_2} \ C^{\nu + q^N}_{\gamma_1 \ \bar{\lambda}_2} \ ,   
\label{GN2}
\end{eqnarray}
where representations $\bar{\lambda}$ are defined in Eqs.(\ref{schur_conj_un}), (\ref{schur_conj_sun}). 
The $N$-ality constraint becomes 
\begin{equation}
|\lambda_1| + |\lambda_2| - |\gamma_1| - |\gamma_2| \ = \ N q \ . 
\label{nality_GN2}
\end{equation}

Three-dimensional case is treated similarly. One finds, {\it e.g.}
\begin{eqnarray}
G_N^3(\lambda_1,\lambda_2,\lambda_3,\gamma_1,\gamma_2,\gamma_3) \ = \ 
\sum_{\nu,\sigma} \ C^{\nu}_{\lambda_1 \ \lambda_2} \ C^{\sigma}_{\gamma_1 \ \gamma_2} \ 
G_N^2(\nu,\lambda_3,\sigma,\gamma_3 ) \nonumber  \\ 
= \sum_q \sum_{\nu,\sigma,\alpha} \ C^{\nu}_{\lambda_1 \ \lambda_2} \ C^{\sigma}_{\gamma_1 \ \gamma_2} \ 
C^{\alpha}_{\nu \ \lambda_3} \ C^{\alpha + q^N}_{\sigma \ \gamma_3} \ . 
\label{GN3}
\end{eqnarray}
The $N$-ality constraint reads 
\begin{equation}
|\lambda_1| + |\lambda_2| + |\lambda_3| - |\gamma_1| - |\gamma_2| - |\gamma_3| \ = \ N q \ . 
\label{nality_GN3}
\end{equation}

\subsection*{\underline{$R_{N,N_f}(r,s;m_f,\mu_f)$}:}

Using Eq.(\ref{trace_power}) for the expansion of the power of traces and Eq.(\ref{fermdet_exp}) for 
the fermion contribution the integral (\ref{Rint_def}) is written as 
\begin{eqnarray} 
R_{N,N_f}(r,s;m_f,\mu_f) &=& \sum_{\lambda \vdash r} \sum_{\nu \vdash s} \sum_{\alpha,\beta} \ 
d(\lambda) d(\nu) \ s_{\alpha^{\prime}}(H_+) s_{\beta^{\prime}}(H_-) \nonumber    \\ 
&\times&\int_G dU \ s_{\lambda}(U) s_{\nu}(U^{\dagger}) \ s_{\alpha}(U) s_{\beta}(U^{\dagger}) \ . 
\label{Rint_gen}
\end{eqnarray}
Integration yields 
\begin{eqnarray} 
R_{N,N_f}(r,s;m_f,\mu_f) = \sum_{q=-\infty}^{\infty}  \sum_{\lambda \vdash r} \sum_{\nu \vdash s} 
\sum_{\alpha,\beta,\sigma} \ C^{\sigma}_{\lambda \ \alpha} C^{\sigma +q^N}_{\nu \ \beta} \ 
d(\lambda) d(\nu) s_{\alpha^{\prime}}(H_+) s_{\beta^{\prime}}(H_-) . 
\label{Rint_res1}
\end{eqnarray}
The last expression can be simplified with help of the formula 
\begin{equation}
\sum_{\mu\vdash r} \ C_{\lambda \ \mu}^{\nu} \ d(\mu) \ = \ d\left ( \nu/\lambda  \right ) \ , 
\label{Cd_sum}
\end{equation} 
where $d\left ( \nu/\lambda  \right )$ is the dimension of a skew representation
defined by a corresponding skew Young diagram. Then, the result of the integration is 
\begin{eqnarray} 
R_{N,N_f}(r,s;m_f,\mu_f) = \sum_{q=-\infty}^{\infty} \sum_{\alpha,\beta,\sigma} \  \delta_{r+|\alpha|,s+|\beta|+qN} 
\nonumber   \\ 
d(\sigma/\alpha) d(\sigma+q^N/\beta) \ s_{\alpha^{\prime}}(H_+) s_{\beta^{\prime}}(H_-) \ . 
\label{Rint_res2}
\end{eqnarray}

\subsection*{\underline{$H_{N,N_f}^d(\lambda_i,\gamma_i;m_f,\mu_f)$}:} 

The case $d=0$ is of special interest as it corresponds to the exactly solvable model of one-dimensional QCD. 
The result of integration can be read off from Eq.(\ref{Rint_res1}) by putting $r=s=0$  
\begin{eqnarray}
H_{N,N_f}^0(0,0;m_f,\mu_f)  &=&  \sum_{q=-\infty}^{\infty} \ \sum_{\alpha,\beta} \  
\delta_{\alpha,\beta + q^N} \ s_{\alpha^{\prime}}(H_+) s_{\beta^{\prime}}(H_-) \nonumber \\ 
&=&\sum_{q=-\infty}^{\infty} \ \sum_{\sigma} \  s_{\sigma}(H_+) s_{N^q \sigma}(H_-) \ . 
\label{Hint_d0}
\end{eqnarray}
The summation over $\sigma$ runs over all partitions such that $\sigma_1\leq N$ and $l(\sigma)\leq N_f$ for the staggered, 
$l(\sigma)\leq 2N_f$ for the Wilson fermions. 
The other cases $d\geq 1$ can be straightforwardly obtained by combining the expansion (\ref{LR_expansion}) with 
the representation (\ref{fermdet_exp}). We find for $d=1$
\begin{eqnarray}
H_{N,N_f}^1(\lambda,\gamma ;m_f,\mu_f)  =  
\sum_{q=-\infty}^{\infty}  \sum_{\alpha,\beta,\sigma} \ C^{\sigma}_{\lambda \ \alpha} C^{\sigma +q^N}_{\gamma \ \beta} \ 
s_{\alpha^{\prime}}(H_+) s_{\beta^{\prime}}(H_-) \ . 
\label{Hint_d1}
\end{eqnarray}
The higher values of $d$ are calculated recursively as 
\begin{eqnarray}
H_{N,N_f}^{d}(\lambda_1,\cdots,\lambda_d,\gamma_1,\cdots,\gamma_d;m_f,\mu_f)  \nonumber   \\ 
=  \sum_{\sigma,\nu} \ C_{\lambda_1 \ \lambda_2}^{\sigma} \ C_{\gamma_1 \ \gamma_2}^{\nu} \
H_{N,N_f}^{d-1}(\sigma,\lambda_3,\cdots,\lambda_d,\nu, \gamma_3,\cdots,\gamma_d;m_f,\mu_f) \ . 
\label{Hint_dn}
\end{eqnarray}

\subsection*{A3: One flavour of staggered fermions}
\label{stag_1fl}

For fixed values of $N_f$ many formulas given above can be specified and simplified by using explicit values 
for the Schur functions which, in turn, can be calculated from Eq.(\ref{ef_hf_schur}). As the simplest but important 
example, let us consider the integral $R_{N,N_f}(r,s;m_f,\mu_f)$ with one flavour of the staggered fermions. 
Taking into account that in this case $\alpha=1^k$, $\beta=1^l$, $0\leq k,l \leq N$ and $s_{\alpha^{\prime}}(H_+)=h_+^k$, 
$s_{\beta^{\prime}}(H_-)=h_-^l$, one gets from (\ref{Rint_res2}) the following simple answer 
\begin{eqnarray} 
R_{N,1}(r,s;m,\mu) = \sum_{q=-\infty}^{\infty} \sum_{k,l=0}^N \sum_{\sigma \vdash r+k}  \ \delta_{r+k,s+l+qN} \ 
d(\sigma/1^k) d(\sigma+q^N/1^l) \ h_+^k h_-^l \ . 
\label{Rint_Nf1}
\end{eqnarray}
Using the similar approach (\ref{Hint_d1}) becomes
\begin{eqnarray}
H_{N,1}^1(\lambda,\gamma ;m,\mu)  =  
\sum_{q=-\infty}^{\infty}  \sum_{k,l=0}^N \sum_{\sigma \vdash |\lambda| + k}
\ C^{\sigma}_{\lambda \ 1^k} C^{\sigma +q^N}_{\gamma \ 1^l} \ h_+^k h_-^l \ .
\label{Hint_d1_Nf1}
\end{eqnarray}
Here the Littlewood-Richardson coefficients can be calculated using the following formula
\begin{equation}
C^{\sigma}_{\lambda \ 1^k} = \begin{cases}
1,\ |\sigma| = |\lambda| + k,\ \lambda_i \leq \sigma_i \leq \lambda_i + 1 , 
\ \lambda_i \geq \sigma_{i + 1} \ ,   \\
0,\ \mathrm{otherwise}\ . 
\end{cases}
\end{equation}

Finally, we specify the result (\ref{Rint_Nf1})  for the physically relevant case, namely $N=3$. For $SU(3)$ the following 
identities hold
\begin{equation}
d(\sigma/1^0) = d(\sigma/1) = d(\sigma) \ \ , \ \ d(\sigma/1^3) = d(\sigma-1^3) 
\label{d_su3}
\end{equation}
that allow us to obtain 
\begin{eqnarray} 
\label{Rint_N3_Nf1}
&&R_{3,1}(r,s;m,\mu) = Q_3(r+1,s) \left ( h_+ + h_-^2 + h_+ h_-^3 + h_+^3 h_-^2  \right )  \\
&+&Q_3(r,s) \left ( 1+ h_+^3 + h_-^3 + h_+^3 h_-^3  \right )  + 
Q_3(r,s+1) \left ( h_- + h_+^2 + h_+^3 h_- + h_+^2 h_-^3  \right )   \nonumber  \\ 
&+&Q_3(r+1,s+1) \left ( h_+ h_- + h_+^2 h_-^2  \right ) + 
Q_3(r+2,s) h_+ h_-^2  + Q_3(r,s+2) h_+^2 h_- \ ,   \nonumber  
\end{eqnarray}
where the function $Q_3(r,s)$ is given in Eq.(\ref{QSUN}). 

\vspace{0.5cm}

{\bf \large Acknowledgements}

\vspace{0.2cm}

O. Borisenko acknowledges support from the National 
Academy of Sciences of Ukraine in frames of priority project 
"Fundamental properties of matter in the relativistic collisions 
of nuclei and in the early Universe" (No. 0120U100935).

\end{document}